\documentclass[twocolumn,tighten]{aastex631}

\newcommand{\ME}{\mbox{$M_{\oplus}$}}
\newcommand{\RE}{\mbox{$R_{\oplus}$}}

\received{October 31, 2024}
\revised{April 21, 2025}
\accepted{April 22, 2025}

\begin{document}

\title{Formation and Evolution Simulations of Saturn, 
Including Composition Gradients and Helium Immiscibility}

\correspondingauthor{Peter Bodenheimer}

\author[0000-0001-6093-3097]{Peter Bodenheimer}
\affiliation{UCO/Lick Observatory, Department of Astronomy and Astrophysics,
University of California, Santa Cruz, CA 95064, USA}
\email{peter@ucolick.org}


\author[0000-0001-9432-7159]{David J. Stevenson}
\affiliation{Division of Geological and Planetary Sciences, Caltech, Pasadena,
CA 91125, USA}
\email{djs@gps.caltech.edu}

\author[0000-0001-6513-1659]{Jack J. Lissauer}
\affiliation{Space Science \& Astrobiology Division, MS 245-3, 
NASA Ames Research Center, Moffett Field, CA 94035, USA}
\email{jack.lissauer@nasa.gov}

\author[0000-0002-2064-0801]{Gennaro D'Angelo}
\affiliation{Theoretical Division, Los Alamos National Laboratory, 
Los Alamos, NM 87545, USA}
\email{gennaro@lanl.gov}




\begin{abstract}
The formation of Saturn is modeled by detailed numerical simulations
according to the core-nucleated accretion scenario. Previous models  
are enhanced to include the dissolution of accreting planetesimals, 
composed of water ice, rock, and iron, in the gaseous envelope of 
the planet, leading to a non-uniform composition with depth. 
The immiscibility of helium in metallic hydrogen layers is also considered.
The calculations start at a mass $0.5$~Earth masses and are extended to 
the present day. 
At 4.57~Gyr, the model, proceeding outwards, has the following structure: 
(i) a central core composed of $100$\% heavy elements and molecules, 
(ii) a region with decreasing heavy element mass fraction, down to a value 
of $0.1$, 
(iii) a layer of uniform composition with the helium mass fraction $Y$ 
enhanced over the primordial value, 
(iv) a helium rain region with a gradient in $Y$, 
(v) an outer convective, adiabatic region with uniform composition in which
$Y$ is reduced from the primordial value, and 
(vi) the very outer layers where cloud condensation of the heavy elements
occurs.
Models of the distribution of heavy elements as a function of radius are
compared  with those derived to fit the observations of the Cassini mission,
with rough qualitative agreement. 
The helium mass fraction in Saturn's outer layers is estimated to be around
$20$\%. Models are found which provide good agreement with Saturn's intrinsic
luminosity and radius.
\end{abstract}



\section{Introduction}\label{sec:intro}
\defcitealias{pollack1996}{P96}

The scientific results from the Cassini space mission have sparked
further interest in Saturn's internal structure, formation, and evolution.
In the past, a number of numerical models have considered this problem.
\citet{pollack1977} published models of the constant-mass cooling
phase, starting at ten times the present radius. They found that at the
present age, the calculated internal luminosity of the planet was a
factor of $3$ lower than the value deduced from observations at the
time. They point out that the immiscibility of neutral helium in
metallic hydrogen could precipitate helium rain-out and provide additional
luminosity, but they did not include it in their calculation. 
A similar conclusion was reached by \citet{fortney2011}.  
The possibility of helium immiscibility was first proposed by 
\citet{smoluchowski1967}, but he thought that convective motions 
would counteract the settling out of the helium. 
Then \citet{salpeter1973} proposed that below some critical temperature,
the helium is insoluble in metallic hydrogen and that the time scale 
for the settling out is shorter than the convective mixing time. 
\citet{stevenson1977} later showed, tentatively, that helium settling 
is necessary to explain the luminosity of Saturn. 

Evolutionary sequences for Saturn, including the effects of helium 
immiscibility, were performed by \citet{hubbard1999}.
They concluded that the value of $Y_1$, the helium mass fraction 
in the outer layers of Saturn, should be between $0.11$ and $0.21$. 
Then \citet{fortney2003} found a phase diagram for hydrogen and helium 
that resulted, after evolution calculations, in agreement with Saturn's 
luminosity and a value for $Y_1=0.185$. However, that phase diagram did 
not result in any helium settling in Jupiter. 

However, $Y_1$ has been measured directly by the Galileo probe in Jupiter 
\citep{vonzahn1998}; the value is $0.236$, slightly depleted from the assumed 
protosolar value of $0.27$. 
The measurement can be used as a calibration point for Saturn calculations, 
but with a residual uncertainty because we do not know the shape of the phase 
diagram (discussed further below). Accordingly, \citet{mankovich2020} made 
calculations for both planets including He immiscibility. The phase diagram for 
H/He that is found to fit the Jupiter measurement for $Y_1$ is then used for 
Saturn models. The models evolve through the cooling phase with a central 
heavy-element core and a H/He envelope in which superadiabatic gradients are 
included in regions with a gradient in the He abundance.
The best-fit model, which agreed with Saturn's luminosity (at the time) at 
the current age, had $Y_1=0.07$, but it required an assumption that 
the Bond albedo in Saturn's atmosphere be $0.50$, rather than the observed 
value, at the time, of $0.342$.
A similar approach was taken by \citet{howard2024}, who assumed three different
phase diagrams of hydrogen plus helium to obtain $Y_1$ at the present time
of $0.16$, $0.13$, and $0.14$.

The observed (highly uncertain) values of $Y_1$ from Voyager and Cassini 
range between $0.07$ and $0.24$ \citep[see][their Figure~14]{mankovich2020}.
For example,
\citet{achterberg2020} found $Y_1\approx 0.08$--$0.14$ from Cassini
measurements. The Cassini results for other planetary properties were 
re-analyzed \citep{wang2024}. 
The estimate of the Bond albedo has increased to $0.41 \pm 0.02$, and that 
of the effective temperature $T_\mathrm{eff}$ has increased from $95.0$ to 
$97.2\,\mathrm{K}$. Together, these changes increase the estimated value 
of Saturn's 
intrinsic luminosity to $\log(L/L_{\odot}) = -9.50$, up from the earlier 
value of $-9.652$.

The calculations mentioned so far included the final evolutionary phase of
Saturn, during which it cools and contracts at constant mass. A calculation 
of  the earlier formation phase 
\citep[case S1 of][hereafter \citetalias{pollack1996}]{pollack1996} 
started with a core of heavy elements with $1$ Earth mass (\ME) and 
considered the accretion of solid planetesimals 
as well as gas from the solar nebula. The solid material added was assumed 
to settle to the central core. The added gas, composed mainly of hydrogen 
and helium, followed an adiabat. The calculations terminated at just above
$40\,\ME$, 
of which $17.5\,\ME$ formed the core, and $23.2\,\ME$ formed the H/He 
envelope. The final accretion phase and the cooling phase were not included. 
The time to reach the end point of the calculation was 
$9.8 \times 10^6\,\mathrm{yr}$. 

The formation time for Jupiter and also for Saturn  can be significantly 
reduced if orbital migration of the forming planet is taken into account 
\citep{alibert2005a,alibert2005b}.
The latter paper, a formation study similar to that of \citetalias{pollack1996},
also includes orbital migration and disk evolution. Jupiter and Saturn accrete
to their current masses in the same disk in times just under $3\,\mathrm{Myr}$; 
both planets migrate inwards.
Including the enrichment of the H/He envelope by dissolving planetesimal 
material can also shorten the formation time for giant planets 
\citep{venturini2016,valletta2020}.
Formation calculations by \citet{lozovsky2017}, for Jupiter, also including 
the enrichment of the H/He envelope by dissolving planetesimals, show that 
composition gradients in the heavy elements develop in the interior, and 
that superadiabatic gradients can result. Their calculations end when the mass 
approaches $1$ Jupiter mass at a time of about $1.7\,\mathrm{Myr}$.
A full calculation for Jupiter, through the formation and cooling phases up 
to the present day, including planetesimal dissolution, is reported by 
\citet{stevenson2022}. 
A recent review of giant planet formation is found in \citet{guillot2023}.

The Jupiter calculations just mentioned were designed to compare the heavy
element distribution with those in models derived from Juno gravity
observations \citep{iess2018}. The various interior models for Jupiter 
and Saturn are summarized by \citet{helled2024} and references therein. 
All models that fit
the data have substantial mixing of heavy elements with light gases.
An example, by \citet{militzer2022}, has a dilute core for Jupiter.
The heavy element mass fraction $Z$ is constant at about $0.2$ in 
a central region, then decreases linearly to about $0.02$ at the half-mass
point, then is constant out to the outer radius. This kind of distribution 
is not consistent with formation calculations
\citep{mueller2020,helled2022,stevenson2022}, 
although these models do show a gradient in $Z$, but a value of $Z=1$ at 
the center.
The disagreement between Jupiter formation models and current gravity-based
interior models is not understood.

The situation in Saturn is somewhat similar in some ways, but very different 
in other ways. The Cassini data from gravity measurements and ring seismology 
indicate that the heavy-element core is dilute, possibly extending out to
$60$\% of the radius \citep{mankovich2021}. The inner region is stable 
against convection and non-adiabatic, with a composition gradient in $Z$. 
Their preferred model has $Z \approx 0.81$ at the center, decreasing to 
$Z \approx 0.02$ at $60$\% of the radius. This is a mere $60$ or so Earth 
masses, very different from the Jupiter models of \citeauthor{militzer2022} 
(dilution out to perhaps $150\,\ME$). The dilutions of Jupiter and Saturn 
must be thought of in terms of mass fraction
involved and not fraction of radius, and are so 
dissimilar that one must consider the possibility that the physical mechanisms 
at play are different. Further models  of the present Saturn are provided by 
\citet{nettelmann2021} based on the gravity data. Models with a compact central
core, a dilute core, or no core can fit the gravity. For example, 
a compact core has $Z=1.0$ out to $15$\% of the radius, then a rather 
rapid decrease to $Z=0.05$ at $30$\% of the radius (see Figure~\ref{fig:7}).

This paper attempts to answer the following questions: 
(1) What is the full history of formation and evolution for Saturn, 
including the heavy-element enrichment of the H/He envelope, starting 
at about $0.5\,\ME$ and ending with the present mass at the present time?
(2) If helium separation in the metallic hydrogen layer is considered, what is 
the likely value of $Y_1$ (mass fraction of He in the outer layers) at 
the present time?
(3) Does the distribution of $Z$ at the present time agree with the results
derived from Cassini observations? The method of calculation, which 
is analogous to that used by \citet{stevenson2022} for Jupiter, is described 
in Section~\ref{sec:2}, the results are reported in Sections~\ref{sec:3} 
and \ref{sec:param}, and concluding remarks are found in Section~\ref{sec:4}.

\section{Method of Calculation}\label{sec:2}

The standard equations of stellar structure and evolution, modified for 
the planetary case, are solved by the Henyey method 
\citep{henyey1964,kippenhahn2013}.
The planet is assumed to be spherically symmetric, non-rotating, and 
in hydrostatic equilibrium, with energy transport by radiation or convection. 
The calculations include the
formation phase, during which the planet accretes solid particles and gas from
the solar nebula, and the following phase of cooling and contraction with no
further accretion, up to the present time. As in the analogous calculations 
for Jupiter \citep{stevenson2022}, it is not assumed that the planet structure
consists of a heavy-element core with uniform composition surrounded by 
an extensive envelope consisting of mainly hydrogen and helium. 
The accreting solid particles dissolve in the H/He envelope, producing 
in some regions gradients in the composition of heavy elements. 
Thus, the energy sources include overall contraction, and gravitational 
energy release from heavy elements settling through the envelope,
and from helium raining out as a result of immiscibility of helium in 
hydrogen during the cooling phase. The main energy sink is radiation from 
the surface. Other sinks include vaporization of the planetesimal material, 
and dissociation and ionization of some of the hydrogen.

\subsection{Planetesimal Deposition}
\label{sec:tesimal}
In the present work, Saturn, forming at $a_p=9.54\,\mathrm{AU}$, where $a$ 
is the distance from the Sun, does not migrate through the protosolar disk.
It accretes planetesimals composed of water ice, rock (silicates), and iron, 
assumed to have mass fractions of $50$\%, $35$\%, and $15$\% of $Z$,
respectively.
The precise composition of planetesimals is not known, and other mass fractions 
are possible.
The accretion rate of solid particles (where $M_Z$ is the total mass, and $Z$ 
the total mass fraction in those particles) is given by 
the standard equation \citep{safronov1969}
\begin{equation}
dM_Z/dt =\pi R_\mathrm{capt}^2 \sigma\Omega F_g,
\label{eq:1} 
\end{equation}
where $R_\mathrm{capt}$ is the capture radius for planetesimals, $\sigma$ 
is the mass per unit area of solid particles in the disk, $\Omega$ is 
the planet's orbital frequency, and $F_g$ is the gravitational enhancement 
factor to the geometrical capture cross section \citep{greenzweig1992}. 
The initial value of $\sigma$ is $3\,\mathrm{g\,cm}^{-2}$. 
Planetesimals, with radius $100\,\mathrm{km}$, enter the gaseous envelope 
at various impact parameters. As accreted solids fall through the
gaseous envelope, the code calculates ablation of material, breakup, and energy 
deposition \citep[\citetalias{pollack1996};][]{gennaro2014}, and determines 
the capture radius of the planet for further solid accretion
\citep{gennaro2014}. 
The integration of the trajectories of the solids within the envelope uses 
a variable-order method with adaptive time-step control to constrain 
the global accuracy \citep{hairer1993}. Breakup occurs occurs when 
the ram pressure $0.5 \rho v^2$ exceeds the compression strength of 
the planetesimal, where $v$ is its velocity relative to the ambient 
medium and $\rho$ is the ambient density. 
The strength is provided by self-gravity as long as the radius of the object 
exceeds a few tens of kilometers \citep[e.g.,][]{gennaro2015}. 
The results from the various impact parameters are averaged to give the added 
mass and energy as a function of radius. This mass sinks all the way to the
central core only very early in the formation process, when the mass of 
the H/He envelope is very low.

For material deposited at a given layer, its partial pressure $P_\mathrm{part}$ 
is compared with the vapor pressure $P_\mathrm{vap}$. 
If $P_\mathrm{part} > P_\mathrm{vap}$, 
the excess vapor above $P_\mathrm{vap}$ sinks (rains out) to a level where 
the temperature allows the two quantities to be equal. For water, silicates, 
and iron, respectively, the vapor pressures in $\mathrm{dyne\,cm}^{-2}$ are
\begin{eqnarray}
P_\mathrm{vap}^\mathrm{H_2O} &=& \exp{(28.867 - 5640.34/T)}        
\label{eq:2}   \\
P_\mathrm{vap}^\mathrm{SiO_2} &=& \exp{(31.302 -54700/T)} 
\label{eq:3}     \\
P_\mathrm{vap}^\mathrm{Fe} &=& \exp{(29.588 -47064.84/T)}
\label{eq:4}
\end{eqnarray} 
where $T$ is the temperature in kelvins at the surface of the solid particle,
for the calculation of ablation, and is the ambient gas temperature for 
the calculation of rain-out.  
Equation~(\ref{eq:2}) is obtained from \citet{podolak1988}, based on data
from the Handbook of Chemistry and Physics \citep{weast1974crc}. 
Equation~(\ref{eq:3}) is based on a numerical fit to data from 
\citet{melosh2007}.
Equation~(\ref{eq:4}) is adapted from the vapor pressure curve for iron
of \citet{podolak1988}.
Above the critical temperature, 
$647\,\mathrm{K}$, $5000\,\mathrm{K}$, and $9500\,\mathrm{K}$
for ice, rock, and iron, respectively,
$P_\mathrm{vap}$ is set to infinity. 
The three species are treated independently. 
We recognize that these vapor pressure relationships and the assumption 
of separated phases is an approximation, especially as one approaches
a critical point (or critical line in the case of mixed phases). 
At present, there is insufficient information on the multi-component phase 
diagrams to enable us to do a better estimate. In particular, 
it is possible that we have overestimated $Z=1.0$ in the very inner regions 
of the planet, since there 
may be significant solubility of hydrogen in silicates or iron as the latter 
approach a supercritical state. This is an open issue for future work. 
(By convention, we consider H in H$_{2}$O as part of $Z$.)

\subsection{Equation of State}
The equation of state of the gaseous envelope is obtained from
\citet{saumon1995} for the very outer regions of the envelope where
the heavy-element abundance is very low. 
There, the mass fraction of hydrogen $X=0.71$, the mass fraction of helium 
$Y=0.273$, and the mass fraction of all other elements $Z =0.017$. 
The  model pressures in the outer layers have been compared with the more 
recent solar-composition table of \citet{chabrier2021} with very good 
agreement.
For hydrogen/helium
gas mixed with rock, iron, and/or water vapor, tables are obtained based
on the quotidian equation of state of \citet{more1988}. These tables
are extended, modified, and compared with other standard equations of state 
by \citet{vazan2013}. The comparisons are with ANEOS \citep{aneos1972}
and SESAME \citep{sesame1992}. Separate tables exist for water
vapor, silicate vapor, and iron vapor. In each case,  sub-tables contain the 
heavy material mixed with various mass fractions of hydrogen/helium
gas, with fractions ranging from $0$ to $1$. If rock, water, and/or iron 
in combination are present at a given layer, the equation of state interpolates 
between the relevant tables, each table weighted by the mass fraction of the
corresponding species. The tables provide the pressure, specific internal 
energy, and adiabatic gradient as functions of density, temperature, 
and composition. 

\subsection{Radiative Opacity}

Convection is the dominant mechanism for energy transport during the formation
and evolution of the planet.
Radiative energy
transport is responsible for the loss of thermal energy within the planet
to space. It can also be important within the planet under
special conditions, such as the outer layers of the planet during the formation
phase, and in regions with a gradient in composition. Thermal conduction has
been shown to be unimportant \citep{hubbard1973}.

The main source of opacity in the outer envelope is dust grains, which  are 
assumed to have a size
distribution ranging from $0.005\,\mu\mathrm{m}$ to $1\,\mathrm{mm}$. 
The grain opacity is exhibited in \citet{gennaro2016}, and the data appear 
in tables calculated by 
\citet[see, e.g., their Figure~14]{gennaro2013}.
Once the grains evaporate, above about $2000\,\mathrm{K}$, the gas opacities 
are taken from \citet{ferguson2005} and \citet{iglesias1996}. 
These sources assume solar abundances of $Z$.
In those regions of the envelope where the composition is $100$\% heavy
elements, a table is  taken from data in the Opacity Project archives 
\citep{seaton1994}. 
The high-metal opacities are high enough so that the regions of the models 
with $100$\% heavy elements are fully convective; therefore the structure is 
insensitive to the opacity values. In the transition region between $100$\% 
heavy elements and solar composition, which encompasses $37$\% 
of the total mass at the end of accretion, opacities are interpolated between 
the solar table and the 
high-$Z$ table. The mass fraction of heavy elements is determined for a 
given zone, and logarithms of the opacities from these two tables are 
interpolated linearly in the mass fraction. 

During the final cooling phases of planetary evolution, after accretion 
stops, the grains are assumed to settle into the interior and to evaporate. 
In those phases almost the entire mass, outside the regions with composition 
gradients, is convectively unstable, except for a thin radiative layer at 
the outer edge. The radiative opacity is dominated by molecular sources 
\citep{freedman2014}. Their Equations~(3), (4), and (5) are used with 
a ratio of hydrogen to metals of $100$.

\subsection{Treatment of Convection}
During the formation phase, the planet typically includes an inner central
core, with uniform composition of 100\% heavy elements and a mass of about 
$1\,\ME$.
It contains material derived from planetesimals during the earliest stages 
of accretion. This core is not modelled in detail, but its mean density is
specified. This density is the maximum of $2.5\,\mathrm{g\,cm}^{-3}$
and that provided by the model at the core's outer boundary.
As accretion continues, an outer core forms still with 100\% heavy elements
($Z=1$). 
It can grow up to $10\,\ME$ and is included in the solution for the entire
planet. 
Farther out, a layer forms where $Z$  decreases sharply with radius. The layer
in which the rock  and iron vapor concentrations decrease and that where 
the water concentration decreases are typically well separated in radius. 
For example, the rock/iron fraction typically decreases in the temperature 
range $3500$--$2500\,\mathrm{K}$ while the water fraction decreases in the 
temperature range $500$--$450\,\mathrm{K}$.
The outer layers have nebular composition (H and He in the solar ratio) out 
to the total radius, except when Helium separation sets in.

The uniform-composition inner and outer cores are unstable to convection 
according to the Schwarzschild criterion \citep[e.g.,][]{kippenhahn2013}, 
and the adiabatic temperature gradient is applied.
The layers with a molecular weight gradient are unstable to convection
according to the Schwarzschild criterion but stable according to the test 
formulated by \citet{stevenson1977} and discussed in \citet{bodenheimer2018}.  
The time scale for  double-diffusive convective mixing of material through
these layers is assumed to be very long compared with the planet formation
time, as suggested by the results of \citet{leconte2012}. Such mixing
would tend to smooth out the composition gradient, and it is neglected here.
Energy transport by convection is suppressed, and radiative transfer 
is included.
Modelling the temperature gradient in this region is uncertain but important, 
because it determines the radiative flux. It is taken to be an average of 
the gradients given by the Schwarzschild and Ledoux criteria 
\citep[e.g.,][]{kippenhahn2013}.
Outside the composition gradient, the composition is again uniform, and if the
layer is unstable to convection according to the Schwarzschild criterion, the
adiabatic gradient is applied.

\subsection{Boundary Conditions}\label{sec:bound}
The inner boundary condition is applied at the mass fraction at the outer edge 
of the inner core, about $1.7\,\ME$. The radius there ($\approx 1.57\,\RE$),
is defined by the inner core mass and its assumed mean density. The luminosity 
is set equal to zero. During Phases~I through III of formation 
\citep[see Section~\ref{sec:form} and Section~1.2 in][]{stevenson2022}, 
the outer boundary condition on the planet assumes a nebular temperature 
of $100\,\mathrm{K}$ and a density of $10^{-11}\,\mathrm{g\,cm}^{-3}$.
The outer radius $R_p$ of the planet is set to 
$R_\mathrm{eff} = R_\mathrm{H}/4$,  
or to $R_\mathrm{eff} = R_\mathrm{B}=GM_p/c^2$, whichever is the smaller. 
Here, $R_\mathrm{H}$ is the planet's Hill radius, and the factor $4$ is 
determined by three-dimensional hydrodynamic simulations \citep{lissauer2009}. 
$R_\mathrm{B}$ is the Bondi radius, $c$ is the sound speed in the disk 
at $a_p$, and $M_p$ is the total planet mass.
The gas accretion rate $\dot{M}_{XY}$ is determined by the requirement that 
$R_p = R_\mathrm{eff}$. 
Note, however, that our definition of $R_\mathrm{eff}$ is approximate, since
its determination should also take into account the local thermodynamics 
of the disk's gas \citep[see, e.g.,][]{gennaro2013,kuwahara2024}.

However, once $R_p$ contracts so rapidly that the disk cannot supply gas fast 
enough to keep $R_p=R_\mathrm{eff}$, disk-limited accretion (Phase~IV) starts. 
$R_p$ rapidly contracts within $R_\mathrm{eff}$ and the planet ``detaches'' 
from the disk. 
The accretion of nebular gas onto the planet is now in nearly free fall. 
The temperature and density at $R_p$ are determined by the procedure of 
\citet{bodenheimer2000}, which approximately takes into account the shock 
that forms at the outer boundary of the hydrostatic planet. Gas and solid 
accretion rates during this phase are calculated according to the method 
described in \citet{gennaro2021}; see also 
\citet[][their Sections~2.5 and 2.6.]{stevenson2022}.
Gas accretion rates are based on numerical estimates obtained from global, 
high-resolution, three-dimensional hydrodynamic calculations of a planet
interacting with a disk \citep{lissauer2009,bodenheimer2013}. 
The calculations span a range of conditions, from marginal tidal gaps 
to deep gaps, and also account for the gas transport far away from the planet.
Therefore, the disk-limited accretion rates applied in the models presented
herein are consistent with both local tidal perturbations in the disk and
larger-scale transport \citep[for a comparison, see][]{tanigawa2016}. 
The parameters for their determination include the unperturbed 
disk surface density at $a_p$, the planet's orbital period, 
the planet mass $M_p$, and the turbulent viscosity parameter $\alpha$ 
of the disk's gas. After disk-limited accretion sets in,
the solid accretion rates are tied to those of gas, as given by Equation~(3) 
of \citet{stevenson2022} with the width parameter $b=4$ \citep{kary1994}.

Once accretion stops, the planet makes a transition to photospheric boundary 
conditions. During the ensuing evolution phase, the planet contracts and cools 
as an isolated object, although heating from the Sun is included in the 
radiative boundary condition according to Equations~(2) through (5) of
\citet{gennaro2016}. 
The equilibrium temperature is set to $79\,\mathrm{K}$, as a result of the
Bond albedo estimate of $0.41$ \citep{wang2024}.  

\subsection{Helium Separation}\label{sec:he}

The H-He phase diagram is not known, though there have been many attempts, 
both theoretical and experimental, to define it. We adopt some general 
principles to guide our model, but acknowledge that considerable uncertainty 
remains. 
The helium rain region is typically defined as being between pressures 
of about $1$ to $2\,\mathrm{Mbar}$. The low pressure cut-off is motivated 
by the experimentally observed rapid increase in free electrons (expressed 
through electrical conductivity) that begins around this pressure. 
These free electrons are expected to interact repulsively with the essentially  
neutral helium atoms, favoring a rapid onset of insolubility as pressure 
increases along a gently changing isentrope. The high pressure behavior 
is poorly understood, but we here assume that the solubility of He is 
eventually roughly constant with pressure (though mildly increasing with
temperature), meaning that the isentrope is steeper than the solubility 
curve and He can then be well mixed (despite being higher mole fraction 
than farther out). Accordingly. our models have a uniform-composition 
He-rich layer, and at lower pressures there is a uniform-composition 
helium-reduced layer. 
The boundaries of the rain region are somewhat adjustable to
ensure conservation of mass. The total mass of helium remains constant, 
that is, in a given time step, the mass of helium subtracted from the
outer region must equal the mass added to the He-rich layer. 
The total masses of hydrogen and the heavy elements ($Z$) are also conserved.

To determine the amount of rain in a given time step, we define the saturation 
helium mole fraction $x_s$
\begin{equation}
   x_s = \exp{(-T_s/T)},
   \label{eq:5}
\end{equation}
in which $T$ is the actual temperature and
\begin{equation}
  T_s = T_1/(1 + \exp{(\beta (P_1-P))},
  \label{eq:6}
\end{equation}
where $T_1$, $P_1$, and $\beta$ are parameters and $P$ is the actual pressure;
both $P$ and $P_1$ are assumed to be in Mbar.
The first of these equations is simply the theoretical prediction for 
solubility arising from balancing the entropic term $k_B \ln{x}$ for 
a mole fraction $x\ll 1$, with the interaction term represented by $k_B T_s$. 
The pressure dependence is not known but chosen so that the onset of 
insolubility with pressure is fast, as argued above.
In practice, in most cases, $P_1$ and $\beta$ are kept constant at 
$1\,\mathrm{Mbar}$ and $10$, respectively, while $T_1$ is varied in 
the different runs.
At the outer edge of the rain region, the helium mole fraction 
is set to $x_s$ (the saturation value) at that point. 
That determines $Y$ throughout the outer adiabatic zone. At the inner edge 
of the rain region, the helium mole fraction is set to $x_s$ at that point.
That determines $Y$ in the inner adiabatic region. The rain region itself, 
with a gradient in $Y$, is not an adiabat. Convection is suppressed; 
radiative energy transport is included.
The defined pressures at the edges of the rain region, as well
as the pressure at the inner edge of the inner adiabatic region,
are adjusted to ensure mass conservation, through reruns, if necessary.
The energy generated by the gravitational settling of He from the outer
to the inner adiabatic zone is included as a contribution to the planet's
intrinsic luminosity.

\begin{figure}[ht]
\centering%
\resizebox{\linewidth}{!}{\includegraphics[clip]{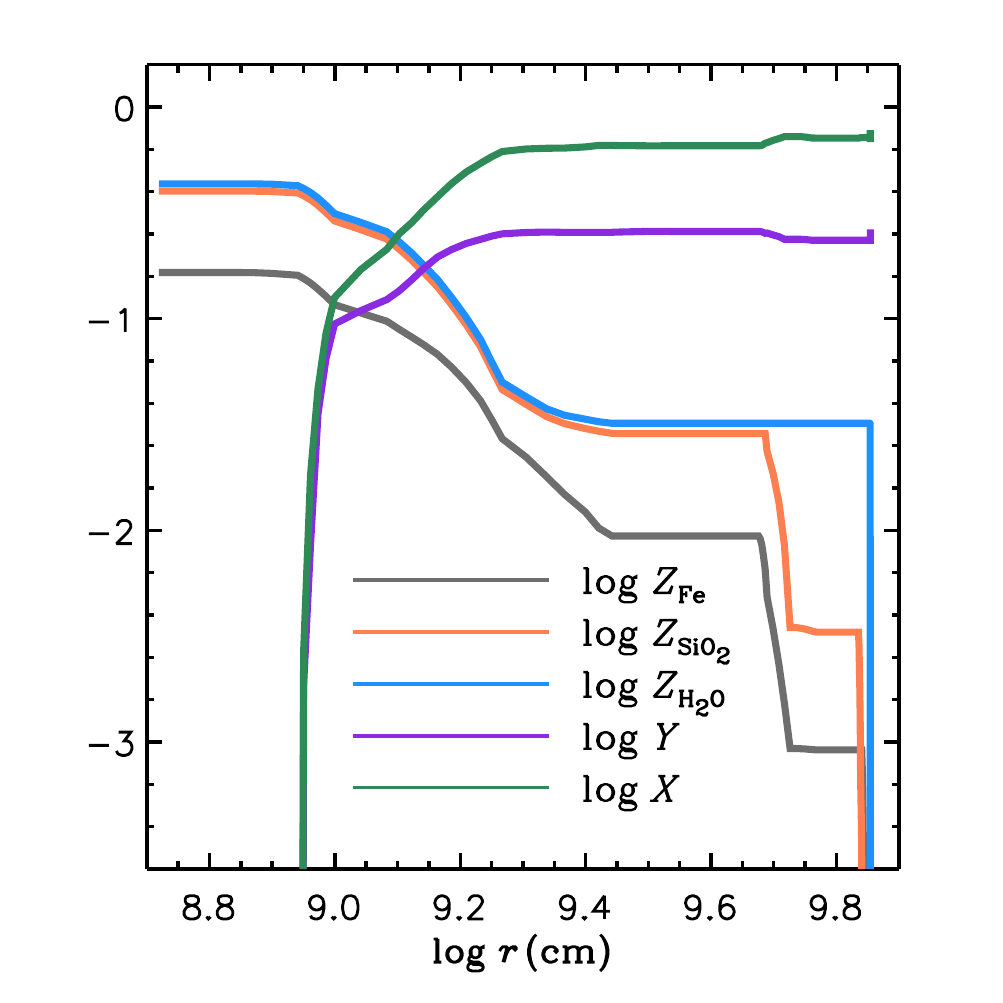}}
\caption{
Distributions of composition (mass fractions) at age $4.57\,\mathrm{Gyr}$
in a model of Jupiter used to calibrate the parametric temperature $T_{1}$
in Equation~(\ref{eq:6}). As a result of He rain-out, the outer layers of
the planet are (slightly) depleted in Helium. 
Quantities, as labeled, are plotted as a function of radius.
}
\label{fig:14}
\end{figure}
The He separation calculations are started approximately $1\,\mathrm{Gyr}$ 
after the beginning of the cooling phase. The value of $T_1$ is determined 
through calculations of the depletion of $Y_1$ in Jupiter models, using 
the same procedure as described in this section. The calibration calculations 
for Jupiter are also started at $1\,\mathrm{Gyr}$ after the beginning of
 the cooling phase, using planet structure models at that epoch obtained 
 from the standard run of
\citet{stevenson2022}. The value of $T_1$ is adjusted through several 
iteration runs until the final value of $Y_1$ at $t=4.57\,\mathrm{Gyr}$ 
reaches $0.236$. 
The deduced value, $T_1 = 15000\,\mathrm{K}$, is applied to the Saturn runs.
For reference, the distribution of composition in the final iteration of 
these Jupiter models is shown in Figure~\ref{fig:14}. 
At time $t=4.57\,\mathrm{Gyr}$,
the calculated radius is $1$\% larger than the actual mean radius of Jupiter,
and the intrinsic luminosity is $\log L/L_{\odot} = -8.99$, compared with 
the observed value of $-8.92$. 
The total heavy-element mass $M_Z$ in the model is $31.3\,\ME$.

\section{Results: Standard Model}\label{sec:3}
\begin{deluxetable*}{lccccccccccch}
\tablecaption{Results from our standard model of formation of Saturn\label{table:smodel}
}
\tablewidth{\textwidth}
\tablehead{
\colhead{} & \colhead{time} & 
\colhead{$M_{c}$} & \colhead{$M_{e}^{Z}$} & \colhead{$M_{e}^{XY}$} & \colhead{$T_{c}$} &
\colhead{$\rho_{c}$} & \colhead{$\log{(L/L_{\sun})}$} & 
\multicolumn{1}{c}{$\dot{M}_{Z}$} & \multicolumn{1}{c}{$\dot{M}_{XY}$} &
\colhead{$R_{p}$} & \colhead{$M_{p}$} & \nocolhead{mod} \\
\colhead{} & \colhead{(Myr)} &
\colhead{(\ME)} & \colhead{(\ME)} & \colhead{(\ME)} & \colhead{($10^{3}\,\mathrm{K}$)} &
\colhead{($\mathrm{g\,cm^{-3}}$)} & \colhead{} &
\multicolumn{2}{c}{($10^{-6}\,\ME\,\mathrm{yr}^{-1}$)} &
\colhead{(\RE)} & \colhead{(\ME)} & \nocolhead{}
}
\startdata
Start & $0.60$ & $0.54$ & $0.00017$ & $1.7\times 10^{-5}$ & $6.780$ & $0.0014$ & $-7.84$ & $1.50$ & $0.19$ & $87.2$ & $0.54$ &
 1725ir3e \\
Max.\ of $M_{c}$& $0.63$ & $1.73$ & $0.57$ & $1.0 \times 10^{-3}$ & $22.5$ & $0.091$ & $-7.31$ & $32.6$ & $0.0326$ & $400$ & $2.30$ &
2600ir4d \\
Max.\ of $\dot{M}_{Z}$ & $0.684$ & $1.73$ & $3.27$ & $9.8 \times 10^{-3}$ & $27.2$ &  $0.116$ & $-6.46$ & $72.0$ & $0.548$ & $895$ & $5.01$ &
6200ir4de \\
End of Phase~I & $0.877$ & $1.73$ & $9.57$ & $0.436$ & $37.7$ & $0.275$ & $-6.91$ & $9.14$ & $9.14$ & $1268$ & $11.74$ &
10300vin 10400ir4op \\
End of Phase~II & $2.69$ & $1.73$ & $14.43$ & $16.16$ & $53.6$ & $0.925$ & $-6.11$ & $16.3$ & $102$ & $1579$ & $32.3$ &
21670ir4op \\
End of Phase~III & $2.79$ & $1.73$ & $16.6$ & $32.5$ & $57.4$ & $1.80$ & $-5.14$ & $59.0$ & $870$ & $1734$ & $50.8$ &
23800ir4opa \\
End of Phase~IV & $2.87$ & $1.73$ & $18.6$ & $74.9$ & $51.8$ & $4.73$ & $-5.94$ & $0.032$ & $0.739$ & $41.8$ & $95.2$ &
35150*cc7 \\
Final model & $4570$ & $1.73$ & $18.6$ & $74.9$ & $22.2$ & $9.19$ & $-9.52$ & $0.00$ & $0.00$ & $9.20$ & $95.2$ &
41504*cc7 \\
\enddata
\tablecomments{
The columns provide the time, the inner core mass ($M_c$), the mass of 
heavy elements in the envelope ($M^{Z}_{e}$), the mass of hydrogen plus 
helium in the  planet, all of which resides in the envelope ($M^{XY}_{e}$), 
the temperature at the inner edge 
of the outer core ($T_c$), the density at the same point ($\rho_c$), 
the radiated luminosity $L$, the solids' accretion rate ($\dot{M}_Z$), 
the accretion rate of hydrogen plus helium from the disk ($\dot{M}_{XY}$), 
the outer radius $R_p$, and the total planet mass $M_p$.}
\end{deluxetable*}

The starting point, 
at time $t=6 \times 10^5\,\mathrm{yr}$,
is a heavy-element core of $0.54\,\ME$ with negligible H/He envelope mass. 
This starting time is an estimate of the time it takes for planetesimals 
to form the initial core 
based on Equation~(\ref{eq:1}).
The characteristics of the planet through all phases of formation and evolution
are summarized in Table~\ref{table:smodel}.
Once the inner core mass $M_{c}$ attains its maximum, accreted solids dissolve 
entirely in the envelope.
By definition, the outer core is the volume surrounding the inner core in which 
$\log Z>-0.01$, i.e., $Z>97.7$\%. 
The densities in these cores vary over time as the planet gains mass and
compresses. 
At an early stage, when the inner core mass is $0.76\,\ME$, the density 
in the inner core is $2.5\,\mathrm{g\,cm}^{-3}$, while that in the outer core 
varies from $3.5 \times 10^{-4}$ to $5.2 \times 10^{-6}\,\mathrm{g\,cm}^{-3}$.
At the conclusion of the run, the inner core density is 
$9.19\,\mathrm{g\,cm}^{-3}$ 
while that in the outer core varies from $9.19$ to $6\,\mathrm{g\,cm}^{-3}$. 
Phase~II ends at the crossover point, when $M_Z=M_{XY}=M_{e}^{XY}$.
Phase~IV ends when gas around the planet's orbit disperses.

\subsection{Formation Phases I--IV}\label{sec:form}
During Phase~I the accretion is dominated by solid particles. 
The inner core (formed by delivery of heavy elements through impacts) increases 
to $1.73\,\ME$ and its mass remains constant thereafter, as planetesimal 
material completely dissolves farther out. The temperature at its outer edge 
increases from $6.78 \times 10^3\,\mathrm{K}$ to $3.8 \times 10^4\,\mathrm{K}$.
The solid accretion rate $\dot{M}_Z$ starts at about 
$1.5 \times 10^{-6}\,\ME\,\mathrm{yr}^{-1}$,
increases to a maximum of $7.2 \times 10^{-5}\,\ME\,\mathrm{yr}^{-1}$,
and then declines again toward the end of the phase.
The radiated luminosity starts at about $\log L/L_{\odot} = -7.84$, increases 
to a maximum of $\log L/L_{\odot} = -6.46$, and declines to 
$\log L/L_{\odot} = -6.9$ at the end.

\begin{figure}[ht]
\centering%
\resizebox{\linewidth}{!}{\includegraphics[clip]{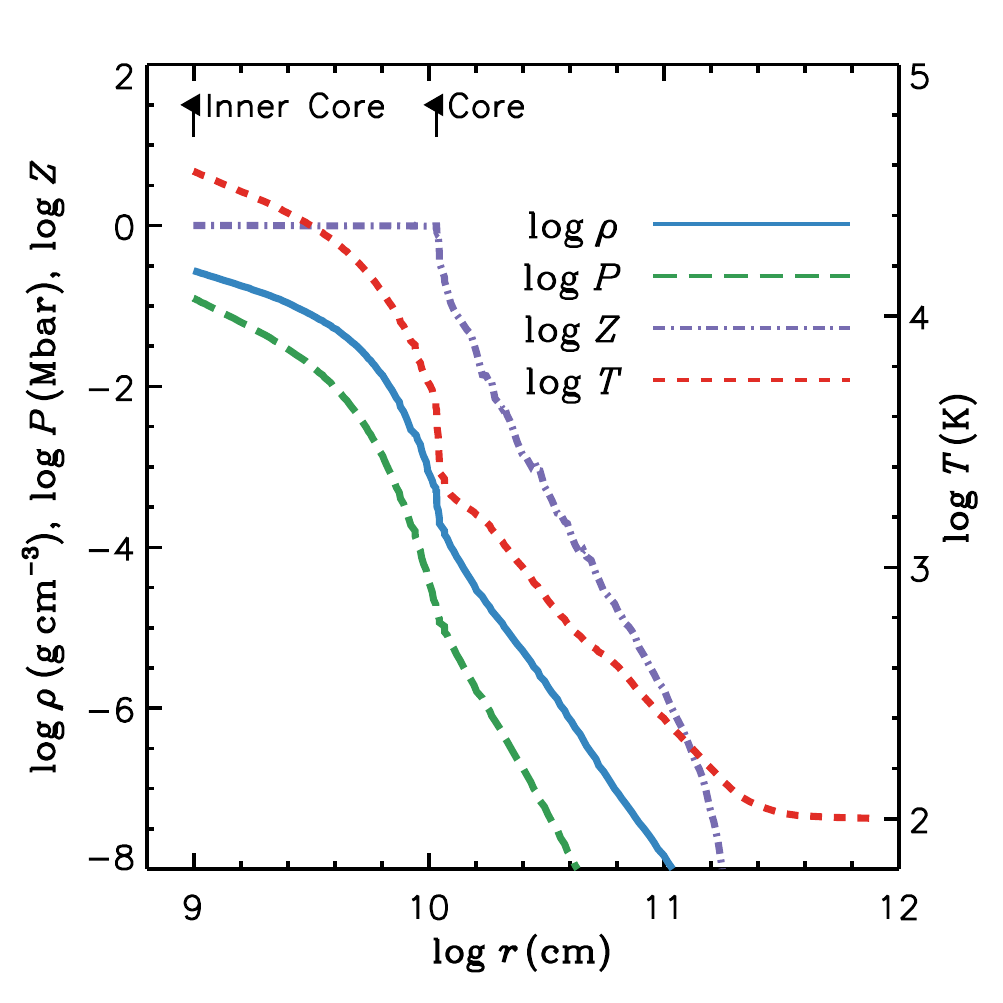}}
\caption{ 
Structure of a model Saturn-mass planet at the end of Phase~I.                
Quantities are plotted as a function of radius, as labeled. The interior
temperature is indicated on the right axis.
The mass fraction of heavy elements, $Z$, includes iron, silicates, and water.
The thin vertical bars near the top left mark the outer boundaries of the inner
and outer cores.
}
\label{fig:1}
\end{figure}

The end of Phase~I is defined by $\dot{M}_{XY}=\dot{M}_Z$, where $Z$ refers 
to heavy-element material while $XY$ refers to hydrogen plus helium. This event 
occurs at $t=8.77 \times 10^5\,\mathrm{yr}$, with $M_Z = 11.73\,\ME$ and
$M_{XY} = 0.436\,\ME$. Almost all of the $Z$ mass is in the inner 
($1.73\,\ME$) and outer ($9.5\,\ME$) cores.
The mass accretion rate is 
$\dot{M}_Z=9.1 \times 10^{-6}\,\ME\,\mathrm{yr}^{-1}$.
The radius of the inner core is $1.57\,\RE$, that of the outer core is 
$17.26\,\RE$, while the total planetary radius is $ 1268\,\RE$. 
The masses in $Z$ and $XY$ are about the same as those in
\citetalias{pollack1996} at the corresponding phase. 
However, $\dot{M}_Z$ is considerably higher in the present case, by a factor $6$. 
The earlier calculation, which assumed a standard core accretion model, has many
differences compared with the present model, and the reason for the discrepancy
in $\dot M_Z$ is unclear.
The structure of the model, showing distributions of density, pressure, and 
temperature, is presented in Figure~\ref{fig:1}. The quantity $Z$ is the sum 
of the mass fractions of water, rock, and iron, which is $1$ at the inner 
edge of the outer core. The mass fractions of Fe and rock 
decline from $0.15$ and $0.35$, respectively, to $0.01$ in the radius range 
$r=16.17\,\RE$ to $r= 17.26\,\RE$, corresponding to a temperature range from
$5000\,\mathrm{K}$ to $2500\,\mathrm{K}$. 
The corresponding  decline in water vapor, from $0.8$ to $0.01$, occurs 
in the radius range $r=16.9\,\RE$ to $r= 26.5\,\RE$  
and a temperature range from $4150\,\mathrm{K}$ to $1500\,\mathrm{K}$.

Earlier, at the maximum of $\dot{M}_Z$, the time is 
$t=6.842 \times 10^5\,\mathrm{yr}$. The radiated luminosity reaches its
maximum slightly later, at $t=6.846 \times 10^5\,\mathrm{yr}$. 
Here, the total $M_Z=5.0$ $\ME$ and $M_{XY} = 10^{-2}$ $\ME$.
Around this time, the rate of energy deposition by incoming planetesimals can
be expressed by $\log L/L_{\odot}= -6.15,$ about a factor $2$ higher than the
luminosity radiated at the surface (see Table~\ref{table:smodel}).  Much of 
the energy deposition occurs just below the layer where the silicate and iron
mass fractions decrease. The suppression of convective energy transport in 
that layer traps some of the deposited heat 
interior to the gradient so that it does not make its way to the surface. 
Later, close to the end of Phase~I, these two quantities are more
nearly equal, since much of the energy deposition occurs exterior to 
the composition gradient. In both cases, there is a gradient in the water
mass fraction farther out, but this layer does not contribute to heat trapping
because energy transport by radiation there is sufficient to carry the emergent
energy flux.                               

During Phase~II, $\dot{M}_{XY}$ typically exceeds $\dot{M}_Z$ by a factor of 
$2$ to $3$, increasing to a factor $6$ at the end. From the end of Phase~I 
to the end of Phase~II, $\dot{M}_Z$ increases from $9.1 \times 10^{-6}$ to 
$1.63 \times 10^{-5}$ $\ME\,\mathrm{yr}^{-1}$. 
The end of Phase~II is defined by $M_Z=M_{XY}$, the crossover point; 
$M_Z=M_\mathrm{cross} = 16.2$ $\ME$. This value is practically the same
as that obtained by \citetalias{pollack1996}, case S1, as the initial solid 
surface density and distance from the Sun were the same in the two cases. 
It also turns out that $M_\mathrm{cross}$ for Jupiter's formation 
\citep{stevenson2022} was the same ($16.1\,\ME$)
because the isolation masses for the two cases were the same. However, 
the times at crossover in the present case and in \citetalias{pollack1996} 
for Saturn were quite different, $2.69 \times 10^6\,\mathrm{yr}$ and 
$9.5 \times 10^6\,\mathrm{yr}$, respectively. The value of $\dot{M}_Z$ at 
this time in \citetalias{pollack1996} was 
$2.7 \times 10^{-6}\,\ME\,\mathrm{yr}^{-1}$, a factor $6$ smaller than 
that in the present calculation. The main differences included the sinking of 
all planetesimal material to the core in the old calculation, differences 
in opacity and equation of state, and the increased capture radius in 
the present case, mentioned above. In calculations by \citetalias{pollack1996}
for Jupiter, a reduction in the grain opacity by a factor $50$ shortened 
the time to the end of Phase~II by about a factor $3$.
Our opacities are in fact lower than those used by \citetalias{pollack1996} 
but not by a factor $50$, rather a factor of roughly $10$. 
The time scale of Phase~II is discussed at length in \citet{pollack1996}.
They found (their Equation~(22)), by a dimensional argument, that the duration
of this phase is $\propto M^{5/3}_{\mathrm{iso}}/L$, where $M_{\mathrm{iso}}$
is the (heavy-element) isolation mass and $L$ is an average of 
the planet's total radiated
luminosity during the phase. The quantity $M_{\mathrm{iso}}$ is the same
for the two cases 
being compared. The quantity $L$ depends on two coupled effects: first, 
the rate of contraction of the planet's gaseous envelope, and second,  
the rate of planetesimal accretion (which depends on the rate of expansion 
of the planet's feeding zone as $M_p$ increases). Our shorter time 
to reach the end of Phase~II is caused by lower grain opacities, which 
increase $L$, thereby allowing the planet to cool, contract and accrete 
nebular gas, and consequently planetesimals, faster.

At crossover, the inner core mass and outer core mass are $1.73$ and
$10.7\,\ME$, respectively.
The corresponding radii are $1.57\,\RE$ and $5.49\,\RE$, while 
the outer radius is $1570\,\RE$. The main region of deposition of planetesimal 
mass, near the breakup point, is $31.39$ to $47.8\,\RE$, well outside the layers
with a gradient in composition. Temperatures there are around $3000\,\mathrm{K}$, 
where all solid constituents can dissolve, but silicates and iron can still
sink if $P_\mathrm{part} > P_\mathrm{vap}$.
The radiated energy at this point is about twice that of the rate of deposition 
of energy by the planetesimals, indicating that overall gravitational 
contraction is contributing significantly to the energy input.

\begin{figure}[ht]
\centering%
\resizebox{\linewidth}{!}{\includegraphics[clip]{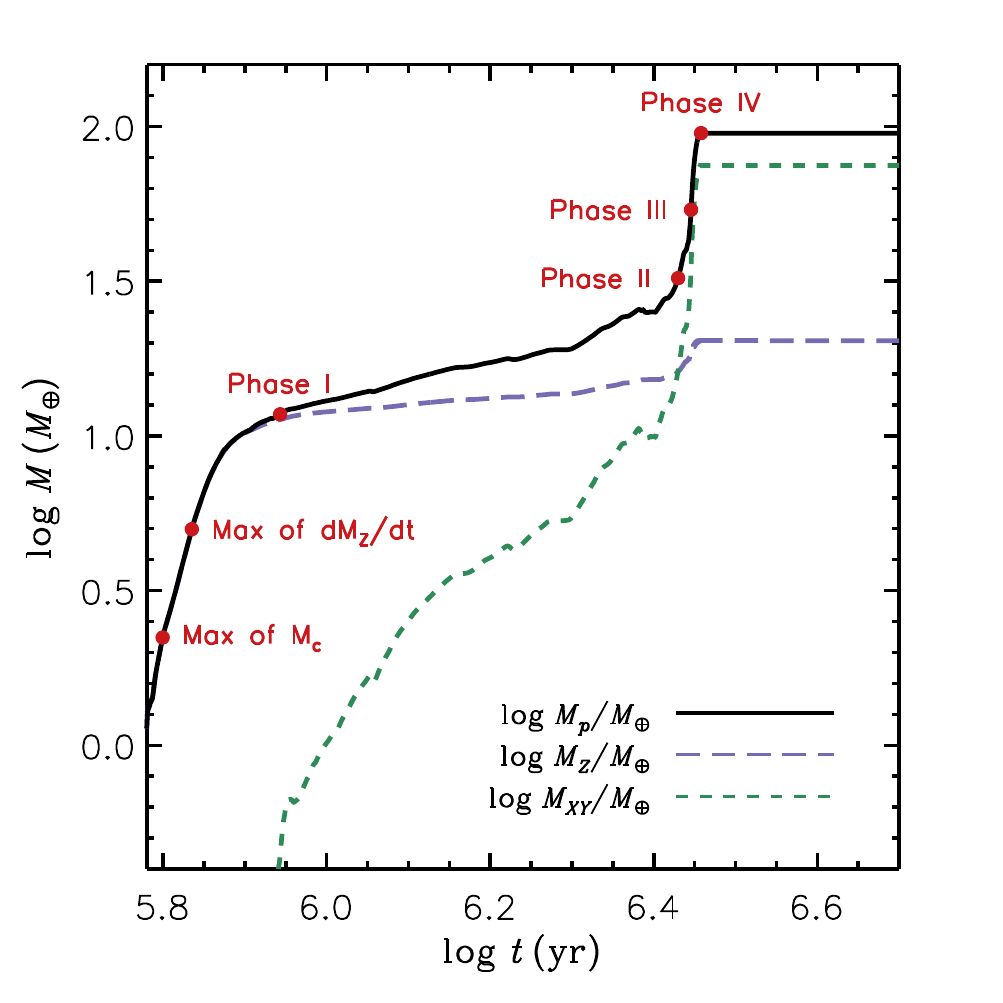}}
\caption{
Evolution of a model Saturn-mass planet up to just past the end of 
accretion.
Time is given in years; masses are in Earth masses, as indicated.
Mass $M_Z$ refers to the total heavy-element mass; mass $M_{XY}$
refers to the mass in hydrogen and helium; the total mass of
the planet is $M_{p}=M_Z+M_{XY}$.
The ends of the various phases are indicated by small circles,
along with the time when the inner core mass, $M_{c}$, stops
increasing and the time when the accretion rate of heavy 
elements, $\dot{M}_{Z}$, attains its maximum value.
Note that some short-term sign variations in $\dot{M}_{XY}$
are likely to be caused by minor numerical inaccuracies in 
the applied accretion procedure (see Section~\ref{sec:bound}). 
}
\label{fig:2}
\end{figure}

\begin{figure}[ht]
\centering%
\resizebox{\linewidth}{!}{\includegraphics[clip]{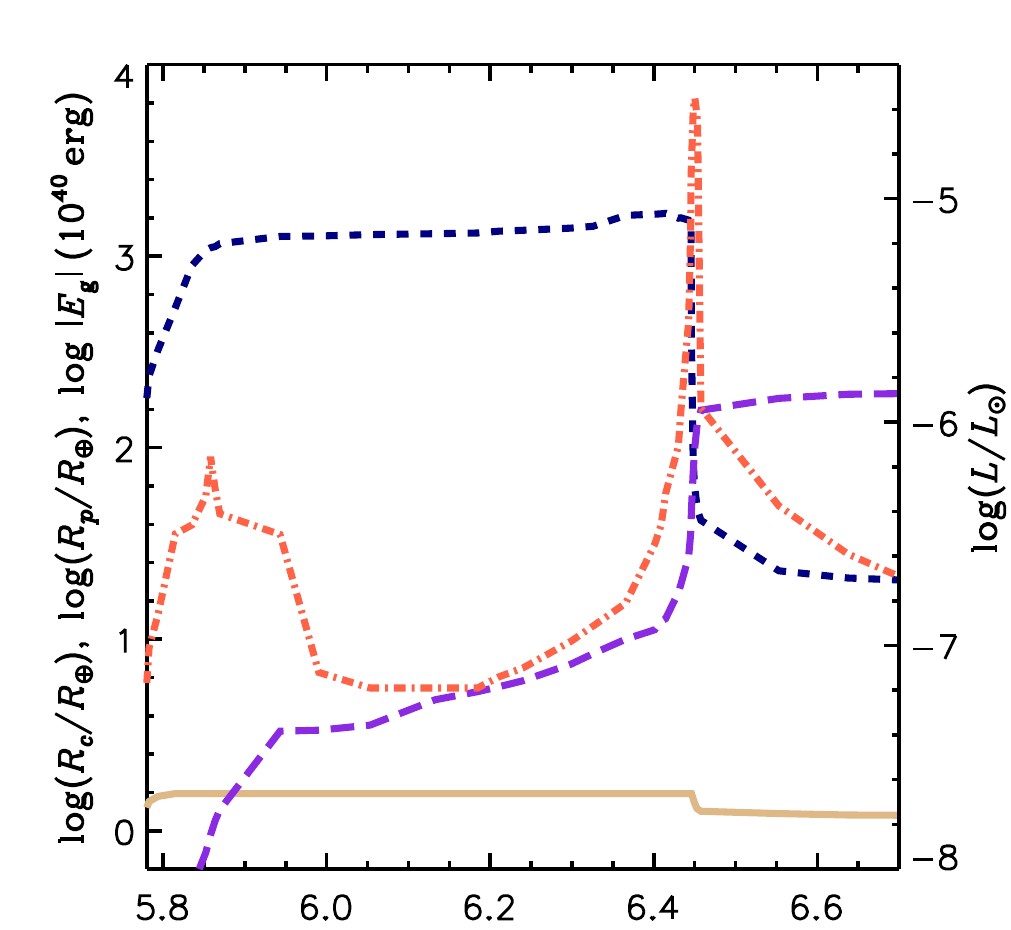}}
\resizebox{\linewidth}{!}{\includegraphics[clip]{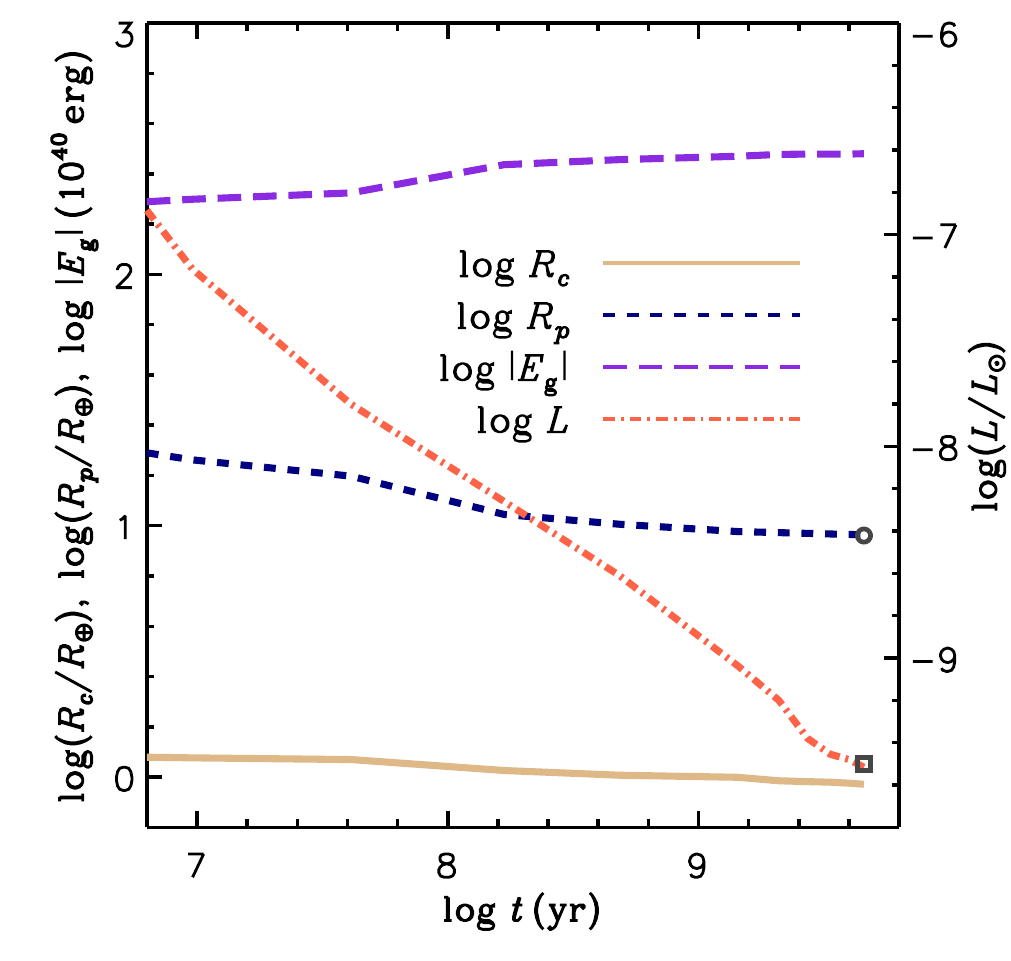}}
\caption{
For the same model as in Figure~\ref{fig:2}, the panels
show: inner core radius $R_{c}$ in Earth radii ($\RE$, 
\textit{solid line});
outer radius $R_{p}$ in units of $\RE$ 
(\textit{short-dashed line}); and the
negative of the gravitational energy $E_{g}$, in units of 
$10^{40}\,\mathrm{erg}$ (\textit{long dashed line}).
The luminosity, $\log{L/L_{\odot}}$ (\textit{dash-dot line}), 
is indicated on the right axes.
Observed values of $R_{p}$ and $L$ are also indicated
in the bottom panel (\textit{open circle and square}).
}
\label{fig:2b}
\end{figure}

Phase~III is characterized by a runaway increase in $\dot{M}_{XY}$ as $M_p$
increases, i.e., $\dot{M}_{XY}/M_p$ becomes larger and larger as accretion
proceeds.
The total mass grows by $18.5\,\ME$, of which $16.3\,\ME$ is $XY$. 
Phase~III ends at $t=2.79 \times 10^6\,\mathrm{yr}$, at which point
the total planet mass $M_p = 50.8\,\ME$ and $M_Z=18.3\,\ME$.
The value of $\dot{M}_Z$ has increased to 
$5.9 \times 10^{-5}\,\ME\,\mathrm{yr}^{-1}$, and $\dot{M}_{XY}$
to $14.7$ times that value.
While $M_p$ has grown by a factor of $1.6$, $\dot{M}_p$ is up by a factor $7.9$.
At this point, planetesimal energy deposition
provides only $1/6$ of the total energy radiated at the surface, with gas
accretion being the dominant source. 
During Phase~III, the planet acquires about $19$\% of its total mass and
$R_p$ reaches its maximum value of $1734\,\RE$.

At $t= 2.79 \times 10^6\,\mathrm{yr}$, the onset of Phase~IV, gravitational
contraction of the outer layers is so fast that the disk cannot provide the
increase in gas mass rapidly enough to keep $R_p=R_\mathrm{eff}$. The  
accretion procedure for Phase~IV is explained in Section~\ref{sec:bound} 
above. Saturn's standard case assumes a disk viscosity parameter 
$\alpha=4 \times 10^{-3}$. 
The accretion rates as a function of $M_p$ are provided in table form 
(and interpolated as necessary).
The accretion rates 
$\dot{M}_{XY}$ and $\dot{M}_Z$ at the beginning and end of Phase~IV are 
given in Table~\ref{table:smodel}.
Throughout Phase~IV, these rates decrease continuously as $M_p$ increases; 
thus the accretion is no longer runaway. 
At the start of this phase, $R_p$ decreases rapidly, by a factor of $10$ 
in a time of $10^4\,\mathrm{yr}$. At the end, the mass contained within 
the inner and outer cores is $12.34\,\ME$, the radius of the inner core 
is $1.26\,\RE$, and the radius of the outer core is $2.67\,\RE$.

\begin{figure}[ht]
\centering%
\resizebox{\linewidth}{!}{\includegraphics[clip]{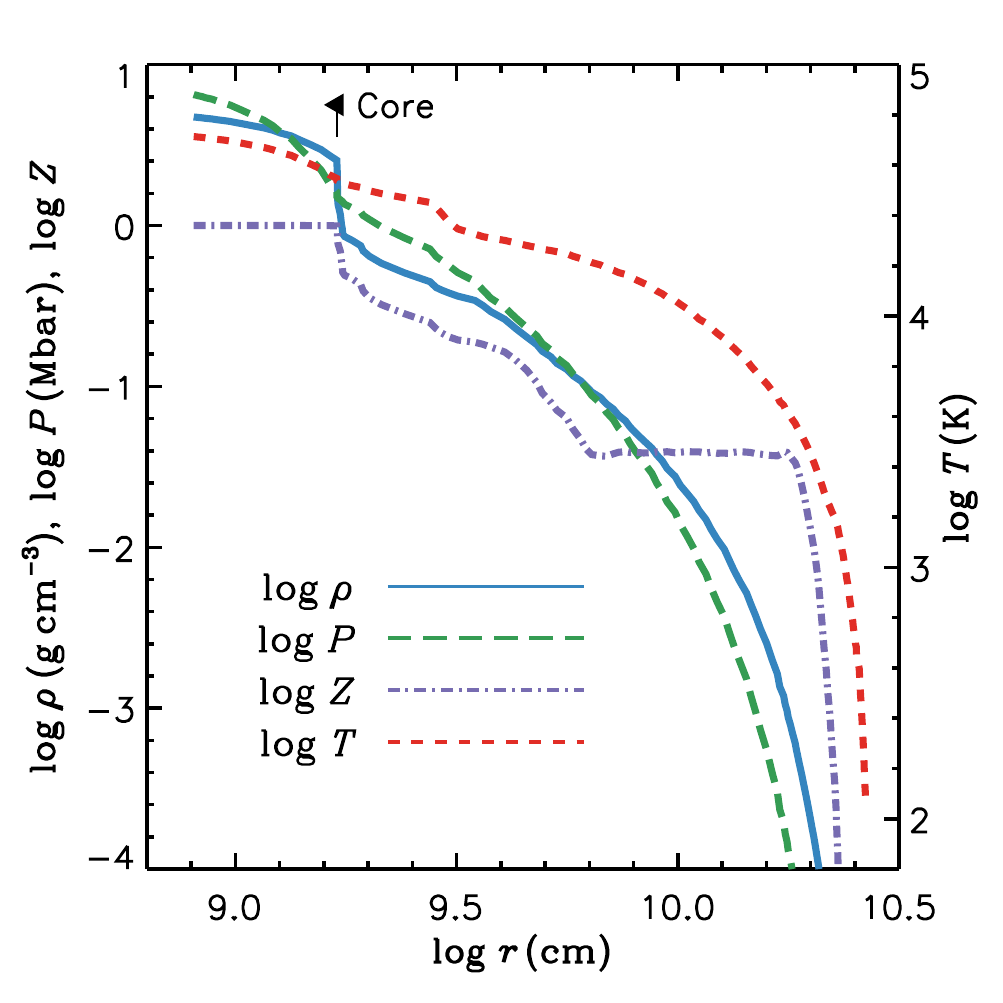}}
\caption{
Structure of a model Saturn-mass planet at age $2.87 \times 10^6$ years,
about the time of the end of accretion. All quantities are plotted as 
a function 
of radius, as labeled. The temperature is indicated on the right axis.
The mass fraction of heavy elements includes iron, silicates, and water.
The total mass of heavy elements is $20.33\,\ME$.
The thin vertical bar at the top marks the outer boundary of the outer core.
}            
\label{fig:3}
\end{figure}

\begin{figure}[ht]
\centering%
\resizebox{\linewidth}{!}{\includegraphics[clip]{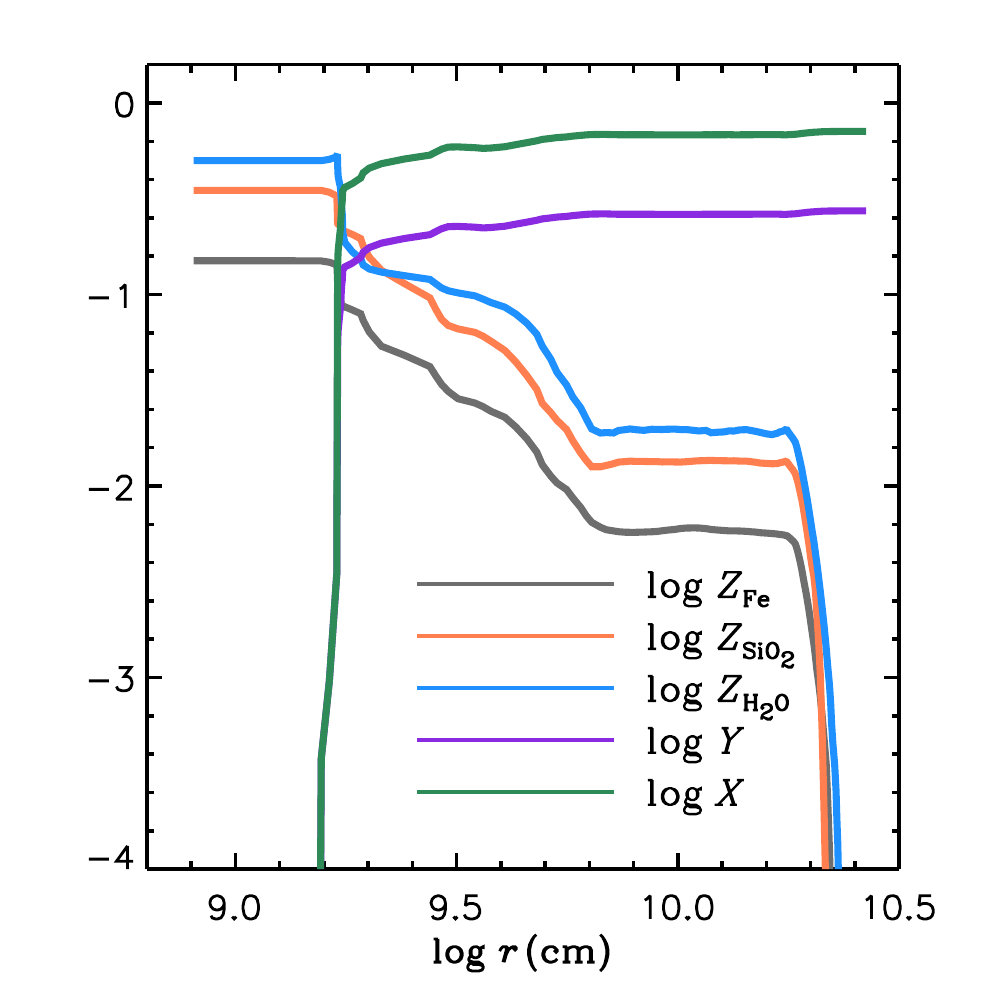}}
\caption{ 
Distributions of composition (mass fractions) in a model 
Saturn-mass planet at the end of accretion 
($t = 2.87 \times 10^6\,\mathrm{yr}$).
Quantities, as labeled, are plotted as a function of radius.
}
\label{fig:4}
\end{figure}

The Saturn mass is reached at about the time of the dissipation of the disk 
($t \approx 2.9 \times 10^6\,\mathrm{yr}$). The evolutions of $M_Z$, $M_{XY}$, 
and $M_p$ are plotted in Figure~\ref{fig:2} up to and just beyond this time. 
In Phase~IV, $M_Z$  increases by $2\,\ME$ to $20.33\,\ME$, 
and $M_{p}$ by $44.4\,\ME$ ($47$\% of the total) to $95.2\,\ME$. 
At this time, the radius $R_p$ decreases to $41.8\,\RE$,
as indicated in Figure~\ref{fig:2b}, which also displays the evolution 
of the planet's luminosity and binding energy, up to the current epoch
(see Figure~\ref{fig:2b}'s caption for further details). 
The first peak in luminosity corresponds closely to the time of maximum
$\dot{M}_Z$, 
while the second peak occurs during the rapid contraction of Phase~IV. 
The structure of the model at the end of Phase~IV is plotted in
Figure~\ref{fig:3}, and the mass fractions of the various species are 
shown in Figure~\ref{fig:4}.

\subsection{Isolated Phase of cooling and contraction}
The calculations of Phase IV show that the gas and solid accretion rates
continuously decrease as the planet mass increases and become negligible 
as the planet approaches 
its final mass of $95.2\,\ME$. The outer boundary condition becomes that of 
a planet isolated from the disk with input of energy from solar radiation. 
The initial He mass fraction in the uniform-composition outer envelope 
is $0.273$. The gradient in heavy-element abundance just outside the outer 
core is maintained, and convection is suppressed in those layers. The outer
envelope is fully convective out to a thin radiative surface zone. 
In the outer layers, the dust grains, whose opacity dominated the formation
phases, are assumed to sink and to evaporate, leaving molecules as the main
source of opacity. 

Starting at a time $t_0=1.43 \times 10^9\,\mathrm{yr}$, He
separation begins. The parameter $T_1$ is set equal to 
$15000\,\mathrm{K}$ (see Section~\ref{sec:he}). 
Somewhat thereafter, at $t=2.09 \times 10^9\,\mathrm{yr}$, the adjusted outer 
edge of the helium rain region is at $P=1.35\,\mathrm{Mbar}$, the inner 
edge at $2.15\,\mathrm{Mbar}$, and inner edge of the inner He-enhanced 
region at $P=2.85\,\mathrm{Mbar}$. The pressure in the region of the 
$Z$-composition gradient at $t_0$ goes from  about $P=9\,\mathrm{Mbar}$ 
to  about $P=1\,\mathrm{Mbar}$. Thus, the rain region and the region of
enhanced He abundance overlap the outer part of the composition gradient 
in $Z$, with $Z$ at $P=2.85\,\mathrm{Mbar}$ equal to $10$\%. It is assumed 
that the buildup of He at the inner edge of the rain region results 
in dynamical instability which mixes He inward, overwhelming the tendency 
of the $Z$-gradient to suppress convection. All chemical species are mixed 
to uniform composition between $P=2.85\,\mathrm{Mbar}$ and 
$P=2.15\,\mathrm{Mbar}$.
These boundaries can change modestly at 
different times or in different runs.
The effects of competing processes on the mixing in this region should be
examined more carefully in the future.

\begin{figure}[ht]
\centering%
\resizebox{\linewidth}{!}{\includegraphics[clip]{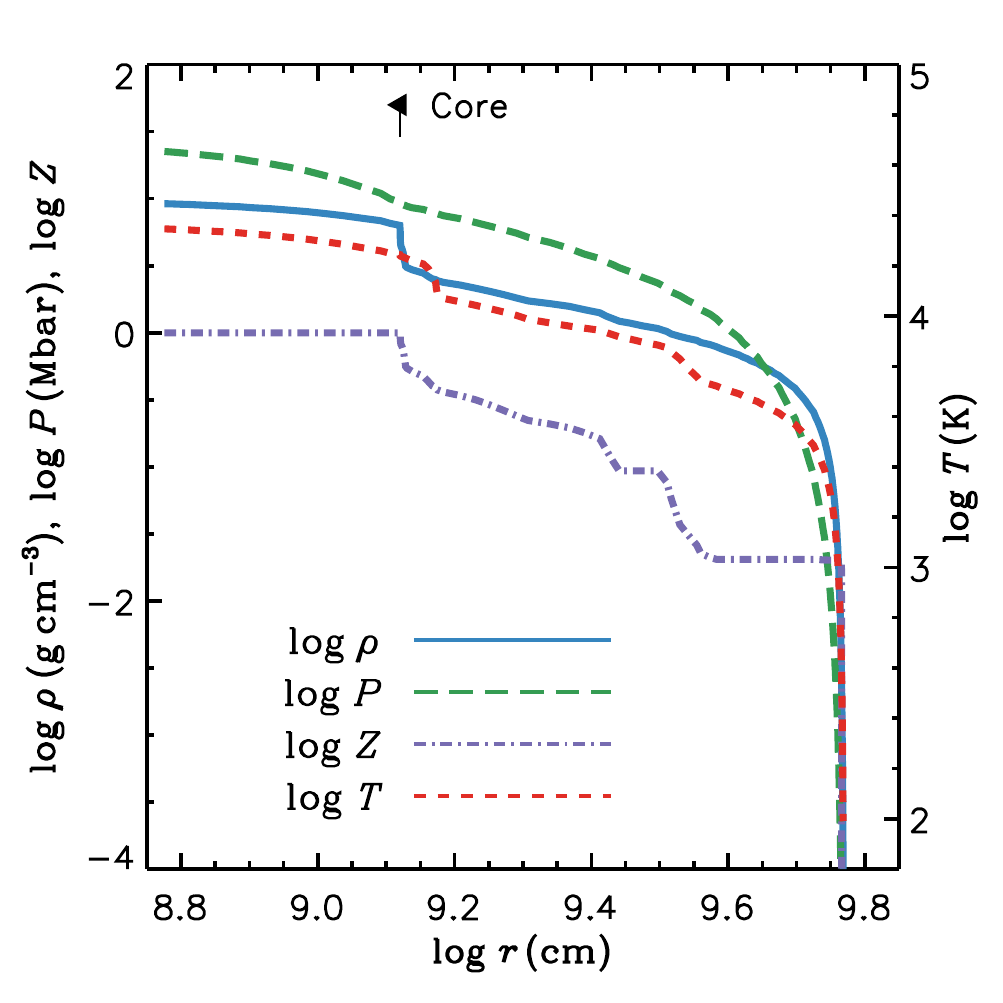}}
\caption{
Structure of a model Saturn-mass planet at age $4.57 \times 10^9$ years.
All  quantities are plotted as a function of radius, as labeled.
The temperature is indicated on the right axis. 
The mass fraction of heavy elements includes iron, silicates, and water.
The thin vertical bar at the top marks the outer boundary of the outer core.
}
\label{fig:5}
\end{figure}

In the final standard model at $t=4.57 \times 10^9\,\mathrm{yr}$, the intrinsic
luminosity $\log L/L_{\odot} = -9.52$ is in good agreement with that measured by
\citet{wang2024}.
The outer radius, $R_p=9.20\,\RE$, is less than $1$\% larger 
than the measured mean value. The model effective temperature is 
$97.7\,\mathrm{K}$, close to the value of $97.2\,\mathrm{K}$
deduced from the observations of \citet{wang2024}. The model temperature at 
$1\,\mathrm{bar}$ pressure is $133\,\mathrm{K}$. The inner and outer boundaries 
of the helium rain region are $P=2.1\,\mathrm{Mbar}$ and $1.47\,\mathrm{Mbar}$, 
respectively. The inner edge of the He-enhanced region is
$P=3.2\,\mathrm{Mbar}$, where $Z=0.093$. 
The mass of the He-enhanced region is $11.2$ $\ME$, and that of the He-depleted
region is $38.4$ $\ME$. The luminosity within the rain region is 
$\log L/L_{\odot}= -11.58$; energy transport is suppressed there because of 
the gradient in the He abundance. In the overlying layers, the value increases 
to $\log L/L_{\odot}= -10.28$.
 
\begin{figure}[ht]
\centering%
\resizebox{\linewidth}{!}{\includegraphics[clip]{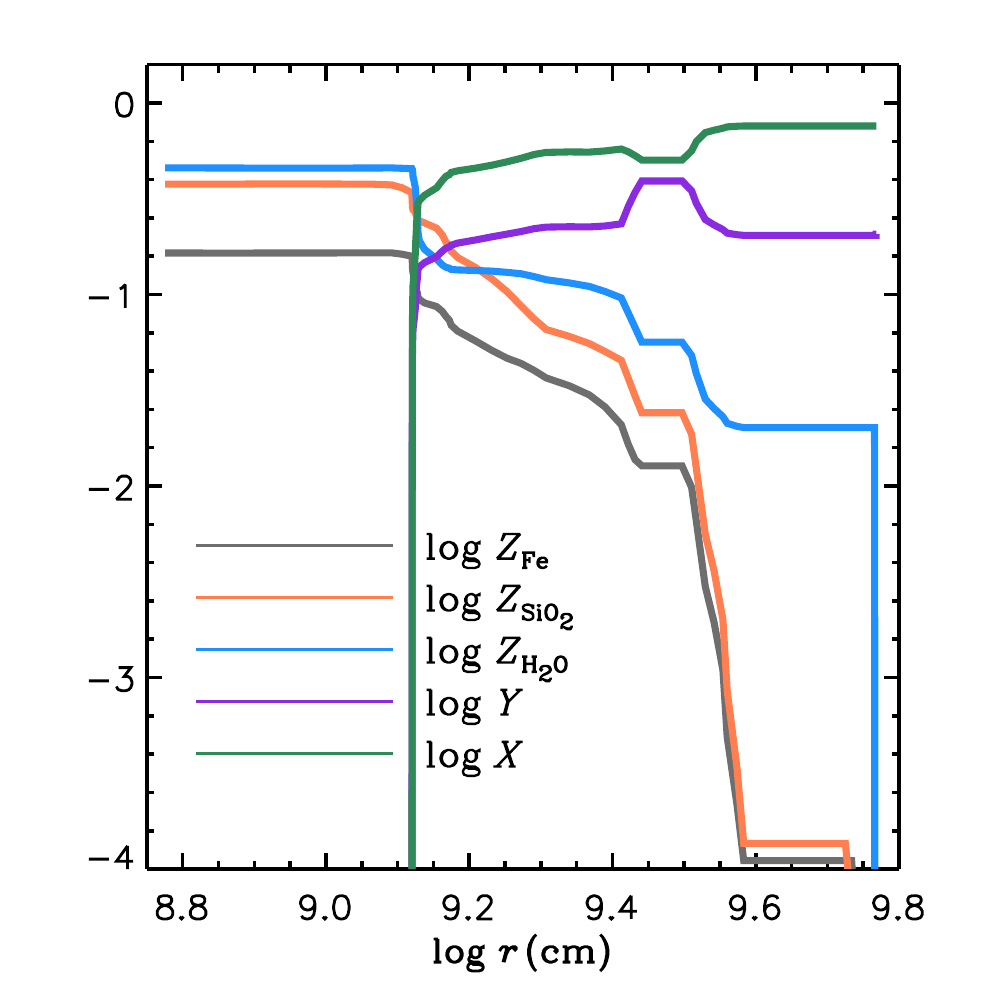}}
\caption{
Distributions of composition (mass fractions) 
in the model Saturn-mass planet shown in Figure~\ref{fig:5}.
Quantities, as labeled, are plotted as a function of radius. 
The sharp drops in the abundances of Fe and SiO$_2$ at about 
$\log r=9.55$ are caused by condensation and rainout. 
The drop in He abundance at this radius corresponds to the rain region.
}
\label{fig:6}
\end{figure}

The value of $Y_1$ is $0.204$. The internal temperature $T_c$, 
at the inner edge of the outer core, is $2.2 \times 10^4\,\mathrm{K}$. 
Two other models were run through the cooling phase with slight changes 
in the boundaries of the helium rain region. The corresponding values 
of $Y_1$ were $0.188$ and $0.215$. All of these values are consistent 
with the observations of Cassini. 
The structure of this final model is shown in Figure~\ref{fig:5}, 
and the composition distributions are shown in Figure~\ref{fig:6}.

\begin{figure}[ht]
\centering%
\resizebox{\linewidth}{!}{\includegraphics[clip]{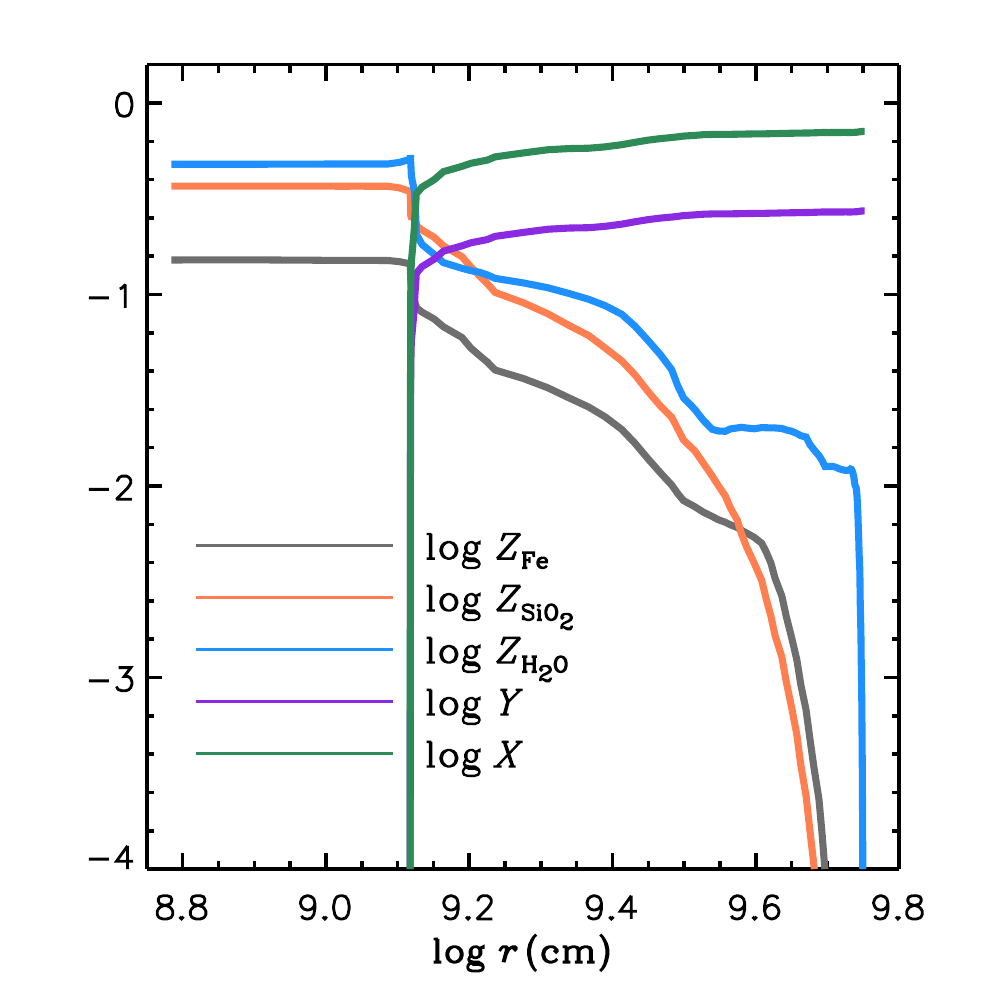}}
\caption{ 
Distributions of composition (mass fractions) in the model Saturn-mass planet 
at age $4.57 \times 10^9\,\mathrm{yr}$, calculated without helium separation. 
Quantities as labelled are plotted as a function of radius. 
Compare with Figure~\ref{fig:6}.}
\label{fig:6a}
\end{figure}

\begin{figure}[ht]
\centering%
\resizebox{\linewidth}{!}{\includegraphics[clip]{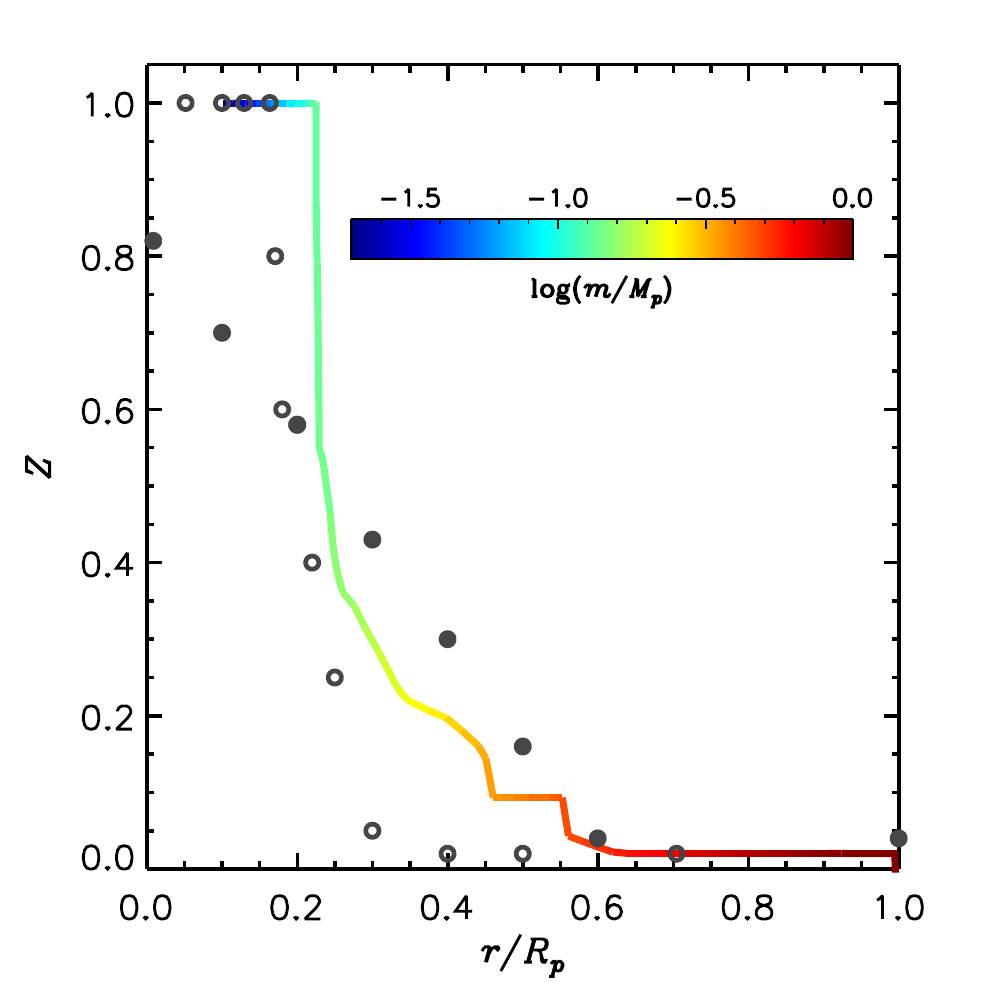}}
\caption{
Mass fraction of heavy elements, $Z$, as a function 
of normalized radius in the model Saturn-mass planet shown 
in Figure~\ref{fig:5}.
The distribution of $Z$ is color-coded by the total mass fraction within 
a given radius. 
The inner core lies inside $r/R_{p}\approx 0.1$.
\textit{Filled circles} represent results of a model from
\citet{mankovich2021} whereas \textit{open circles} are results from 
a centrally condensed model of \citet{nettelmann2021}. 
The radius in these latter two models is normalized by the actual radius 
of Saturn.
}
\label{fig:7}
\end{figure}

\begin{figure}[ht]
\centering%
\resizebox{\linewidth}{!}{\includegraphics[clip]{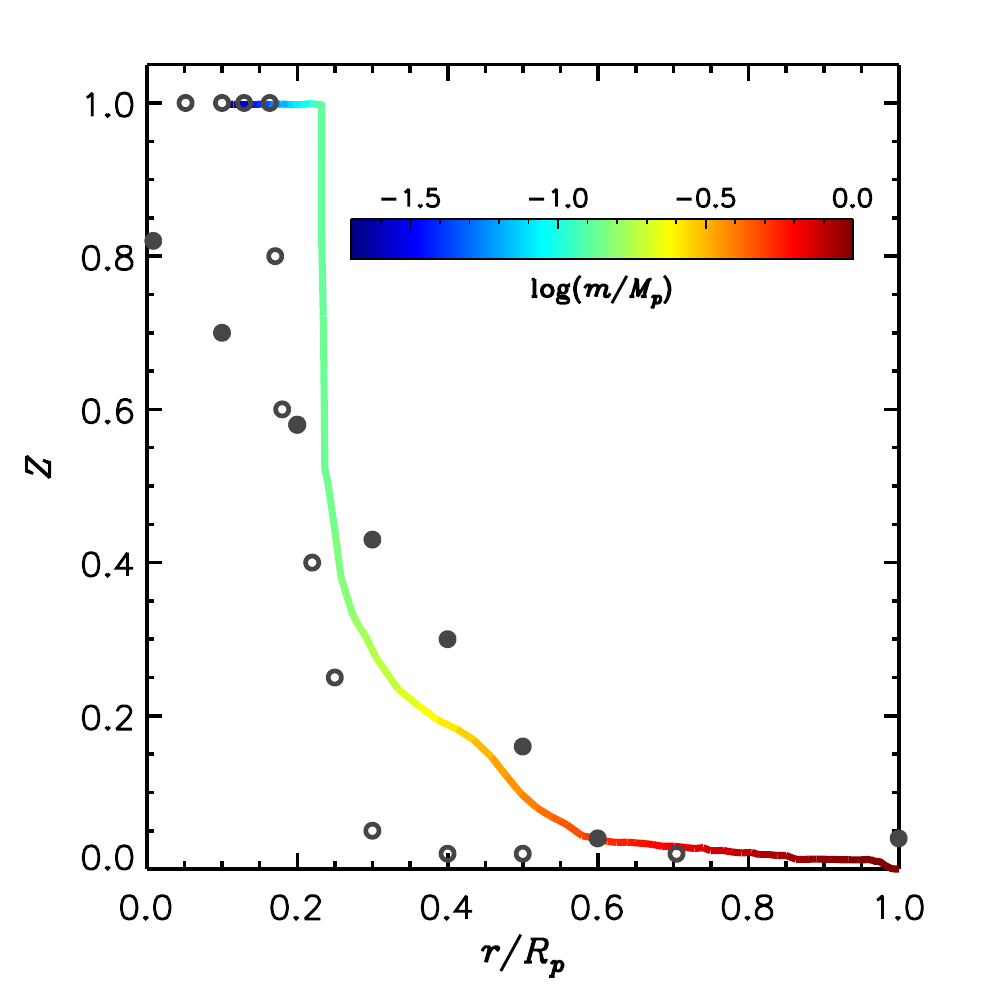}}
\caption{ 
Mass fraction of heavy elements, $Z$, as a function of 
normalized radius in the model Saturn-mass planet shown in Figure~\ref{fig:6a}
(model without helium separation).
The distribution of $Z$ is color-coded by the total mass fraction
within a given radius. 
The inner core lies within $r/R_{p}\approx 0.1$.
\textit{Filled circles} represent results of a model from \citet{mankovich2021};
\textit{open circles} are results from a centrally condensed model
of \citet{nettelmann2021}.
}
\label{fig:8}
\end{figure}

For comparison purposes, a model was computed through the cooling/contraction
phase without the consideration of helium separation. The resulting
composition distribution is shown in Figure~\ref{fig:6a}. The agreement with 
current observed values is not as good as in the standard case. The model 
intrinsic luminosity is $\log L/L_{\odot} = -9.74$, just over half the observed 
value. The effective temperature is $93\,\mathrm{K}$, lower than 
the standard value of $97.2\,\mathrm{K}$.
The computed radius is $8.87\,\RE$, about $3$\% below the observed value. 
The discrepancy in luminosity is smaller than the factor of $3$ obtained 
in the past. The present models, however, have significant differences 
from those earlier ones \citep[e.g.,][]{pollack1977}. Here, the deposition 
of heavy elements in the envelope results in higher internal temperatures 
and a steeper average temperature gradient. 
The composition gradient in $Z$ results in some super-adiabatic layers. 
Note that the earlier models assumed an adiabatic H/He envelope surrounding 
a heavy-element core. Also, the present calculation includes the accretion 
phases, including disk-limited accretion, and therefore provides a different
initial condition for the onset of the cooling phase than was assumed in 
the earlier cases, which considered only that late phase. 
In any case, the calculation with helium separation provides a better fit 
to observations of the present Saturn than does that without separation.
The $Z$ distribution in the model with He separation at 
$t=4.57 \times 10^9\,\mathrm{yr}$ is shown in Figure~\ref{fig:7}. 
It is qualitatively similar to that shown in Figure~\ref{fig:8}, without
helium separation.
 
Models of the present Saturn based on observations from Cassini suggest that 
the planet has a diffuse core. The models of \citet{nettelmann2021} are based
on gravity measurements. They admit a variety of possible solutions; 
the model with a centrally condensed core is represented in Figure~\ref{fig:7}. 
Other possible models have central heavy-element abundances $Z=0.6$ to $0.7$.
The total amount of heavy elements in \citeauthor{nettelmann2021}'s models is
in the range $12.6$--$13.6\,\ME$.
The models of \citet{mankovich2021} are based on data from both gravity
measurements and ring seismology. The most likely one according to these 
authors is also represented in Figure~\ref{fig:7}. 
It has central $Z=0.8$ with a linear decline in $Z$ out to about $60$\% of
Saturn's radius. 
Models with higher central $Z$ are possible, but they are less likely. 
The total amount of heavy elements in \citeauthor{mankovich2021}'s models is 
$19.1\pm 1\,\ME$.
The standard model from our calculations has a central region (inner and outer
core) with $Z > 0.98$ out to $22.5$\% percent of $R_p$ and with a mass of 
$12.3\,\ME$. The outer part of the region with a $Z$-gradient has been replaced
by the inner He-enhanced adiabat, in which $Z$ is uniform at $0.093$.
Thus, $Z$ falls rapidly with radius, reaching $Z=0.02$ (the outer envelope 
value) at $60$\% percent of the radius. If the entire $20.3\,\ME$ of heavy 
elements in our model were concentrated in a central core with the same density
as obtained for our inner core, at $4.57 \times 10^9\,\mathrm{yr}$, the core
radius would be about $25$\% of the planet radius. 
Thus, our model with enrichment in $Z$ in the envelope has resulted in a more
dilute core than in standard core-envelope planet models.

\section{Alternatives to the Standard Model}\label{sec:param}

The principal results presented above are for a single simulation 
of Saturn's accretion and evolution, which we henceforth refer to 
as the ``Standard'' case. In order to test the sensitivity of our 
results to various parameters/assumptions, we first perform four different 
simulations, each of which uses the results from the Standard simulation 
for the early growth of the planet, but which diverges from this case 
at later times. Later, we perform two new simulations that start from the 
initial time.
\subsection{Late-stage modifications}
\label{sec:late}
The first of these four simulations is the standard case but without
helium rain, described in the previous section. The following paragraphs
discuss the remaining three. 
The end results for all the calculated cases that reach the present age 
are given in Table~\ref{table:modelrecap}.

\begin{figure}[ht]
\begin{center}
\resizebox{\linewidth}{!}{\includegraphics[clip]{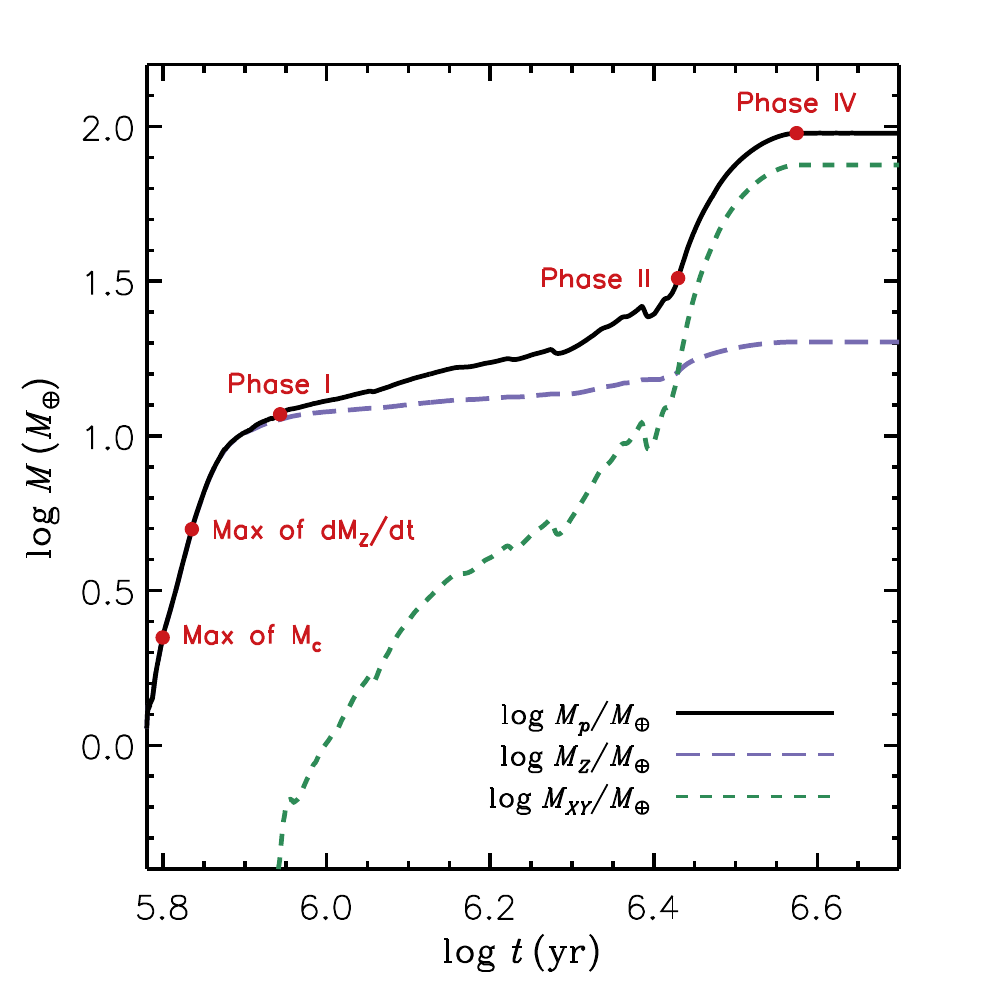}}
\caption{ 
Evolution of a model Saturn-mass planet up to just past the end 
of accretion in the VISC run (see section~\ref{sec:late} for details).
Time is given in years; masses are in Earth masses, as indicated.
Symbols are the same as in Figure~\ref{fig:2}.
The ends of the various phases are indicated by small circles. 
Phase~IV begins at the end of Phase~II, and there is no Phase~III.
Note that short-term decreases in $M_{XY}$ are likely to be caused by 
minor numerical inaccuracies in the accretion procedure.}
\label{fig:15}
\end{center}
\end{figure} 

\paragraph{Viscosity of the Protoplanetary Disk}
This simulation (VISC), that deviates from the Standard run at crossover,
emulates a lower value of the viscosity parameter $\alpha$ in 
the gaseous protoplanetary disk during Phase~IV of disk-limited accretion.
In the standard case (see Table~\ref{table:smodel}), with 
$\alpha=4 \times 10^{-3}$, Phase~IV lasted only $8 \times 10^4\,\mathrm{yr}$
and resulted in an increase of $M_p$ from $50.8$ to $95.2\,\ME$. 
In Run VISC, disk-limited accretion assumes a value of the parameter 
$\alpha \approx 4 \times 10^{-4}$, which produces a deeper gap along the orbit
of the planet and lowers the overall gas transport through the disk, thereby
resulting in lower accretion rates, a prolonged disk lifetime, and
a significant increase in the planet's formation time. 

Run VISC  starts at crossover in the standard case, at 
$t=2.69 \times 10^6\,\mathrm{yr}$ and $M_p = 32.3\,\ME$. 
Phase~III is eliminated, because the condition that the disk supply enough 
material to keep $R_p=R_\mathrm{eff}$ already fails just after crossover. 
The maximum value of $\dot{M}_{XY}$, which occurs at the beginning of Phase~IV, 
is $1.02 \times 10^{-4}\,\ME\,\mathrm{yr}^{-1}$. In the standard case, 
the corresponding value is $8.7 \times 10^{-4}\,\ME\,\mathrm{yr}^{-1}$. 
The maximum radiated luminosity in Phase~IV of the VISC run is 
$\log L/L_{\odot}= -5.32$ at $t=2.9 \times 10^6\,\mathrm{yr}$
and $M_p=76.6\,\ME$. The corresponding quantities for the standard case are 
$\log L/L_{\odot} = -4.57$ at $t=2.53 \times 10^6\,\mathrm{yr}$ and
$M_p=79.9\,\ME$. 
Phase~IV ends at $t=3.466 \times 10^6\,\mathrm{yr}$ with radius 
$R_p=24.2\,\RE$. 
The value of $M_Z=20.0\,\ME$ is slightly less than that in the standard case 
because of differing solid accretion rates in Phase~IV 
\citep[see][]{stevenson2022}.  
The total time spent in Phase~IV is $1.07 \times 10^6\,\mathrm{yr}$, more than 
ten times that of the standard case. Mass versus time during formation, 
and up to this point, is shown in Figure~\ref{fig:15}.

Cooling, contraction, and He separation continue as in the standard case. 
The phase diagram parameters are the same. 
At $t=4.57 \times 10^9\,\mathrm{yr}$, 
the value of $Y_1=0.214$. The intrinsic luminosity is in excellent agreement 
with the observed value, and $T_\mathrm{eff}=98\,\mathrm{K}$ is close to 
the actual value of $97.2\,\mathrm{K}$.
The mean radius is $R_p=9.25\,\RE$, about $1.2$\% larger 
than the actual value. As in the standard case, the temperature at the  
boundary between the inner and outer cores is $T_c= 2.16 \times 10^4\,\mathrm{K}$.

\paragraph{Phase~V added to the standard case}
In Run V, solid accretion continues after the termination of gas accretion
at the end of Phase~IV. During this ``Phase~V'', only heavy elements are
accreted.
The standard case is modified after that time by applying an accretion rate 
of heavy elements $\dot{M}_Z = 10^{-7}\,\ME\,\mathrm{yr}^{-1}$, lasting 
about $2.2 \times 10^7$ years. The added solid material is mixed through the
outer convection zone. 
For this run, we adjusted gas dissipation parameters near the end of
Phase~IV so that gas accretion stopped when the planet was $2.2\,\ME$ short of
Saturn's total mass, enforcing the correct total mass of the planet after Phase~V.
The value of $Z$ in the outer part of the planet 
($2000$ to $200\,\mathrm{K}$) at the end of Phase~V is $0.063$. 
The total mass of heavy elements in the model is $M_Z= 22.55\,\ME$.
That is, about $2.2\,\ME$ of heavy elements were added during Phase~V.  

Cooling and contraction then continue as normal. Helium separation sets in 
at $t=2.1 \times 10^9\,\mathrm{yr}$. The parameters $T_1$, $P_1$, and $\beta$ 
are the same as in the standard case. At $t=4.57 \times 10^9\,\mathrm{yr}$,
He rain-out produces the value $Y_1= 0.229$. The agreement with the observed 
intrinsic luminosity at that time is not as good as in the standard case, 
about $15$\% too high. The value of the effective temperature, 
$T_\mathrm{eff} = 100\,\mathrm{K}$, is about $3\,\mathrm{K}$ too high. 
The mean radius has about the same accuracy as in the standard case. 
However, the added heavy element mass in Run V results
in the calculated radius falling about $0.7$\% below the observed value.
In the standard case, the calculated value is about $0.7$\% too high. 
The final interior temperature, $T_c$, is about $1.3 \times 10^4\,\mathrm{K}$. 
An alternate run with a higher final $T_c = 4.5 \times 10^4\,\mathrm{K}$ 
produced almost the same value of $Y_1$ as quoted above. The latter run
adjusted the parameterized temperature gradient in the
region with a composition gradient in $Z$ to suppress cooling.

\paragraph{Changed parameters for the phase diagram}
All calculations so far have assumed that the parameters in the formula 
describing the phase diagram (Equations~\ref{eq:5} and \ref{eq:6}) are 
$T_1=15000\,\mathrm{K}$,
$P_1=1\,\mathrm{Mbar}$, and $\beta=10$. An alternate calculation (P2) is
carried out, during the phase of helium separation, in which the parameter 
$P_1$ is set to $2\,\mathrm{Mbar}$.
A new calibration is required, so the standard Jupiter model with helium
separation is re-computed with $P_1=2\,\mathrm{Mbar}$, iterating over the 
parameter $T_1$, to determine the value that results in $Y_1=0.236$ for 
Jupiter at the current age. 
The best value found was $T_1=18500\,\mathrm{K}$.

In Run P2, the typical helium rain region falls between $2$ and
$3\,\mathrm{Mbar}$, 
whereas in the standard case it is between $1$ and $2\,\mathrm{Mbar}$. 
The region of enhanced helium concentration (here $3$ to $4.5\,\mathrm{Mbar}$) 
again overlaps with the $Z$-gradient region set up during the formation phases, 
from $9\,\mathrm{Mbar}$ down to about $1\,\mathrm{Mbar}$. In the Jupiter runs, 
this overlap does not occur, as the main $Z$-gradient region falls between $40$ 
and $20\,\mathrm{Mbar}$. 
At the present epoch, Run P2 results in a value $Y_1=0.202$, not significantly 
different from the value obtained from the standard case, despite the difference
in the parameters applied to the phase diagram. The calculated values of radius 
and intrinsic luminosity differ by $\approx 0.5$\%  from the observed values.
The extent of the $Z$-gradient region predicts a diluted core out to $54$\% of 
the planet radius.
The interior temperature, at the outer boundary of the inner core, is 
$T_c = 2.1 \times 10^4\,\mathrm{K}$.

\subsection{Test calculations starting from the initial time} \label{sec:tests}

\begin{figure*}[ht]
\begin{center}
\resizebox{\linewidth}{!}{\includegraphics[clip]{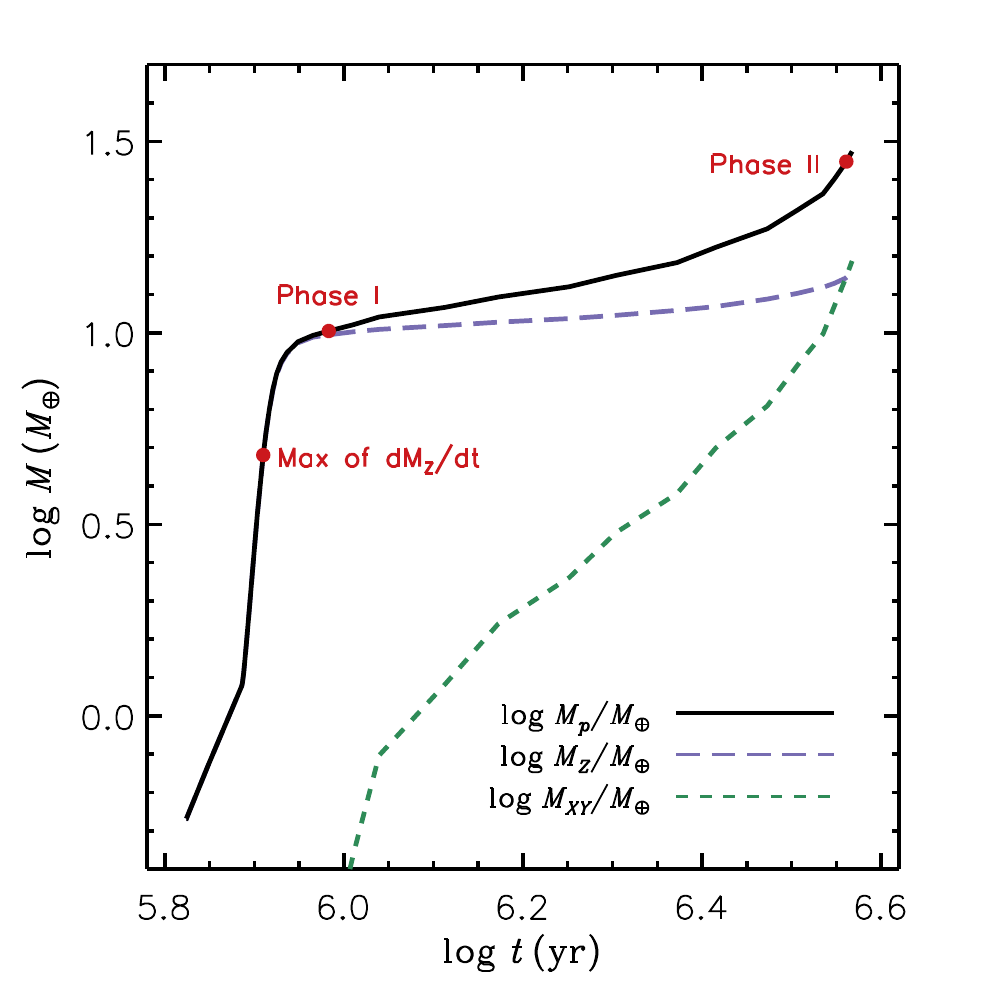}\includegraphics[clip]{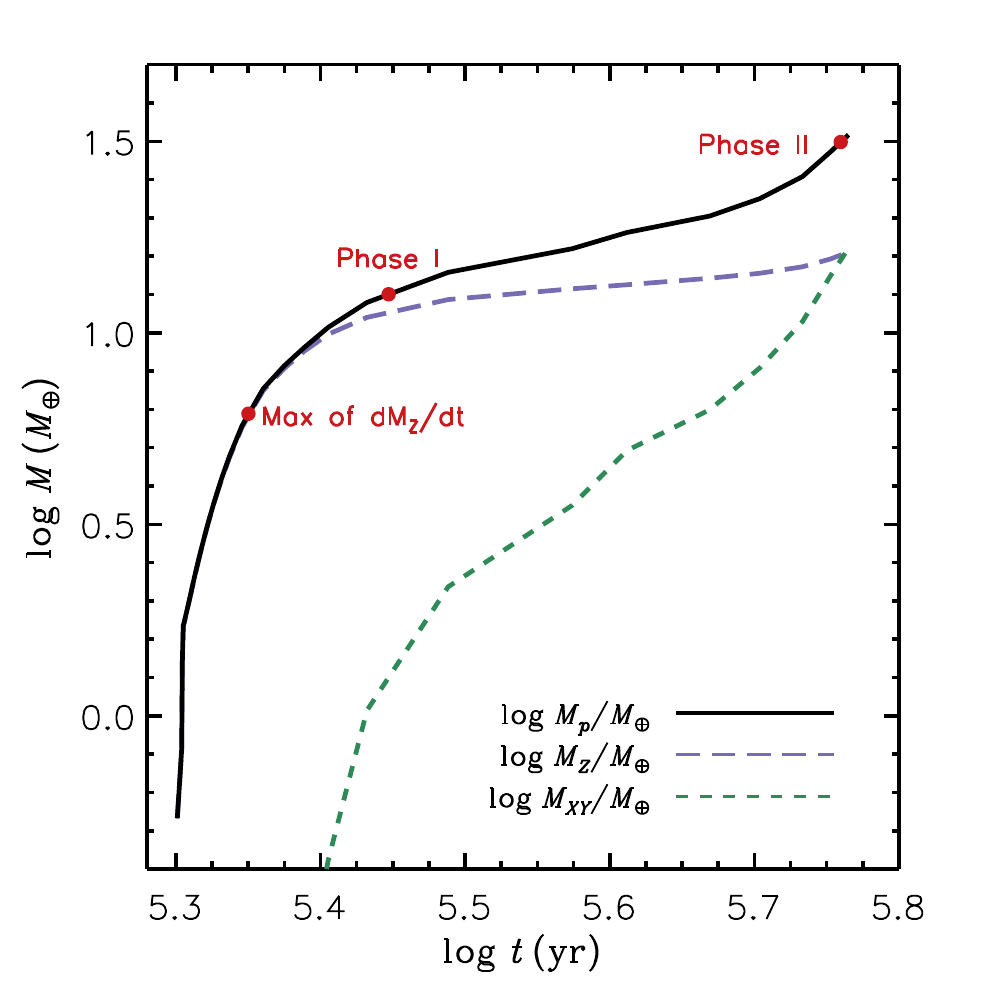}}
\caption{ 
Evolution of Saturn models, up to the end of
Phase~II, for test cases with an initial surface density of solids 
$\sigma =2.7\,\mathrm{g\,cm}^{-2}$ (left) and with smaller planetesimals
of $5\,\mathrm{km}$ in radius (right). See section~\ref{sec:tests} for details.
Time is given in years; masses are in Earth masses, as indicated.
Symbols are the same as in Figures~\ref{fig:2} and \ref{fig:15}..
}
\label{fig:13ab}
\end{center}
\end{figure*}

Numerous parameters and assumptions are involved in these simulations. 
Here we test two of them: the solid surface density $\sigma$, and 
the planetesimal size. The purpose of the calculations is to determine 
the effects on the formation time and the $M_Z$ in the final planet, 
but not to compare the results with the properties of the current planet.
Thus, these calculations only simulated the planet's accretionary era, 
up through Phase~II.

\paragraph{Reduced initial solid surface density}.
The quantity $\sigma$, set to $3\,\mathrm{g\,cm}^{-2}$ in the standard
calculation, has a direct effect on the mass accretion rate $\dot{M}_Z$. 
Here we rerun the calculation \textit{ab initio} with 
$\sigma =2\,\mathrm{g\,cm}^{-2}$.

The calculation starts at a core mass of $0.54\,\ME$, as in the standard case,
and with an envelope mass of $2.7 \times 10^{-4}\,\ME$ of which about $10$\% 
is Hydrogen/Helium. The estimate for the starting time is 
$9 \times 10^5\,\mathrm{yr}$. The starting value of 
$\dot{M}_Z=9.8 \times 10^{-7}$ $\ME\,\mathrm{yr}^{-1}$, compared with 
$1.48 \times 10^{-6}$ $\ME\,\mathrm{yr}^{-1}$ in the standard case, 
a difference of a factor 1.5. Maximum $\dot{M}_Z$ is reached at time 
$t=1.35  \times 10^6\,\mathrm{yr}$, with 
$\dot{M}_Z = 4.74 \times 10^{-5}\,\ME\,\mathrm{yr}^{-1}$. 
In the standard case, the corresponding time is 
$6.84 \times 10^5\,\mathrm{yr}$. The total mass at this point is 
$3.1\,\ME$, of which $2 \times 10^{-3}\,\ME$ is H/He. 
At this time, $\log L/L_{\odot} = -6.53$, slightly less than that 
in the standard case.

The value of $\dot{M}_Z$ steadily decreases, down to 
$4.16 \times 10^{-8}\,\ME\,\mathrm{yr}^{-1}$ at 
$t=1.79  \times 10^6\,\mathrm{yr}$ and $M_p=6.81\,\ME$. At about this time, 
the gas accretion rate first exceeds the solid accretion rate, which defines
the transition to Phase~II. The heavy-element mass is 6.51 $\ME$ and the H/He
mass is $0.3\,\ME$.

Phase~II is followed up to a time $t = 1.05 \times 10^7\,\mathrm{yr}$, at which
point $M_Z = 10.2\,\ME$ and the hydrogen-helium $M_{XY} = 8.5\,\ME$, close to
crossover. The solid accretion rate 
$\dot{M}_Z = 8.9 \times 10^{-7}\,\ME\,\mathrm{yr}^{-1}$, compared with 
$\dot{M}_Z=1.6 \times 10^{-5}\,\ME\,\mathrm{yr}^{-1}$ in the standard case.
The H/He accretion rate $\dot{M}_{XY}$, which dominates during this phase, is
tied to $\dot{M}_Z$, as explained in \citet{pollack1996}. 
Phase~II is about $4$ times as long as in the standard case. 
Typically, in Phase~II, 
as compared with the standard case, $R_{\mathrm{capt}}$ is somewhat smaller,
$F_g$ is comparable, but $\sigma$ is smaller by a factor $3$ or more. 
It is clear that a Saturn-mass
planet cannot form before the dissipation time of the disk, which for the
Jupiter case with similar physics \citep{stevenson2022} is 
$3.14 \times 10^6\,\mathrm{yr}$. 
At this time in the present calculation, the total mass is $M_p= 8.1\,\ME$.

A further test is carried out with the initial value of 
$\sigma=2.7\,\mathrm{g\,cm}^{-2}$. The initial conditions are the same as 
in the standard case, and the starting time is set to 
$6.67 \times 10^5\,\mathrm{yr}$.
Maximum $\dot{M}_Z = 1.18 \times 10^{-4}\,\ME\,\mathrm{yr}^{-1}$ at 
$t=8.13 \times 10^5\,\mathrm{yr}$. The end of Phase~I occurs when
$M_p=10.35\,\ME$, at $t=9.62  \times 10^5\,\mathrm{yr}$. 
The duration of Phase~I is $3.3 \times 10^5\,\mathrm{yr}$, compared with 
$2.77 \times 10^5\,\mathrm{yr}$ in the standard case. Phase~II ends with
a crossover mass $M_{XY} = M_Z = 13.93\,\ME$ at 
$t=3.64 \times 10^6\,\mathrm{yr}$, compared with $16.16$ $\ME$ at 
$t=2.69 \times 10^6\,\mathrm{yr}$ in the standard case. 
The evolution of the model, up to this point, is plotted in the left panel 
of Figure~\ref{fig:13ab}.
Assuming that 
the lengths of the short Phases~III and IV are about the same in the two 
models, the $\sigma=2.7\,\mathrm{g\,cm}^{-2}$ case ends with a formation
time of $ 3.82 \times 10^6\,\mathrm{yr}$ and with $M_Z= 18.1\,\ME$, compared 
with $t=2.87 \times 10^6\,\mathrm{yr}$ and $M_Z= 20.33\,\ME$ in the standard 
model.
This and previous calculations
\citep{pollack1996} indicate that the outcome is very sensitive to the assumed
value of $\sigma$.

\paragraph{Calculations with a smaller planetesimal size}. 
The planetesimal size has a significant effect on the solid accretion rate, 
mainly because of changes in the velocity dispersion and the accretion cross
section. A second calculation is performed {\it ab initio} with planetesimals
of radius $5\,\mathrm{km}$, as compared with the standard value of 
$100\,\mathrm{km}$.
All other parameters are the same as in the standard case. The estimated time 
to reach the starting mass of $0.54\,\ME$ is $2 \times 10^5\,\mathrm{yr}$, 
shorter than that in the standard case mainly because of the increased 
capture cross section. At the start, the value of $\dot{M}_Z$ is roughly 
five times as high as that in the standard case. Maximum $\dot{M}_Z$ occurs 
at $t=2.2 \times 10^5\,\mathrm{yr}$ with a value of 
$2.26 \times 10^{-4}\,\ME\,\mathrm{yr}^{-1}$, 
a factor of $3.1$ as high as that in the standard case. 
The total mass $M_p$ at this point is $5.7\,\ME$, compared with $5.0\,\ME$
in the standard case. Also $M_{XY} = 8.0 \times 10^{-2}\,\ME$, compared 
with $1.0 \times 10^{-2}\,\ME$ in the standard case. The radiated luminosity
$\log L/L_{\odot}= -5.99$, about a factor $3$ as large as that in the standard
model.

The end of Phase~I, where $\dot{M}_Z = \dot{M}_{XY}$,
occurs after an elapsed time of $8.0 \times 10^4\,\mathrm{yr}$ after the start, 
a factor of $3.5$ as fast as that in the standard case. The value 
$\dot{M}_Z= 3.8 \times 10^{-5}\,\ME\,\mathrm{yr}^{-1}$ is a factor $4$ as high as
 the standard value, and $\log L/L_{\odot} = -6.126$ is a factor of 
$6$ as high as the standard value. The actual time is 
$t=2.80 \times 10^5\,\mathrm{yr}$, at which point $M_Z=11.4\,\ME$, 
and $M_{XY}=1.11\,\ME$, $2.5$ times as high as the standard value.

Phase~II runs until $t=5.75 \times 10^5\,\mathrm{yr}$, when the calculation
reaches crossover, with $M_Z = M_{XY} = 15.97\,\ME$, effectively the same 
as in the standard case. The elapsed time of Phase~II is 
$2.95 \times 10^5\,\mathrm{yr}$, only $16.3\%$ as long as that in the standard 
case. At crossover, $\dot{M}_Z=4.055 \times 10^{-5}\,\ME\,\mathrm{yr}^{-1}$ 
(a factor of $2.5$ as high as that in the standard case), while $\dot{M}_{XY}$ 
is  $5$ times as high as $\dot{M}_Z$. The calculation ends just past
crossover.
The mass history of the planet, up to this point, is shown in the right panel 
of Figure~\ref{fig:13ab}.

Phases~III and IV are expected to be very similar to those in the standard
case, because those phases are dominated by disk physics rather than by 
the properties of planetesimals. Thus we assume that those phases last 
a total of $1.8 \times 10^5\,\mathrm{yr}$ (see Table~\ref{table:smodel}). 
The estimated formation time to reach a mass $M_p= 95.2\,\ME$ for the case 
of small planetesimals is $7.55 \times 10^5\,\mathrm{yr}$, far shorter than 
the expected disk lifetime of a few million years.

In this regard, a more realistic situation should include a size distribution 
of planetesimals, which evolves via collisions, gravitational interactions and
aerodynamic drag. Such a scenario was considered by 
\citet{gennaro2014,gennaro2021} for the formation of Jupiter. 
They modeled an evolving distribution of planetesimals
with a minimum size of $15$ meters. The bulk of the planet's heavy elements 
was provided by bodies larger than tens of kilometers in size. 
In terms of formation time, outcomes were comparable to those of
\citet{stevenson2022}, which used the same solids accretion parameterization 
applied in the standard case presented herein.

\section{Conclusions}\label{sec:4}

\begin{deluxetable*}{lccccch}
\tablecaption{Summary of results from Saturn models at the current age\label{table:modelrecap}
}
\tablewidth{\textwidth}
\tablehead{
\colhead{} & \colhead{Mean Radius} & \colhead{Intrinsic luminosity} & \colhead{Effective Temperature} &
\colhead{$1\,\mathrm{bar}$ Temperature} & \colhead{$Y_{1}$} & \nocolhead{mod} \\
\colhead{} & \colhead{($10^{9}\,\mathrm{cm}$)}   & \colhead{$\log{(L/L_{\sun})}$} & \colhead{($\mathrm{K}$)}   &
\colhead{($\mathrm{K}$)} & \nocolhead{} & \nocolhead{}
}
\startdata
Measured  & $5.823$ &  $-9.500$  & $97.2$ &  $134$  &  Unknown &             \\
Standard  & $5.865$ &  $-9.516$  & $98$   &  $133$  &  $0.204$ &  32780*yx1  \\
No He Rain& $5.665$ &  $-9.751$  & $92$   &  $111$  &  $0.273$ &  32770*nohe \\
Run P2    & $5.863$ &  $-9.532$  & $97$   &  $134$  &  $0.202$ &  32780*mhe2 \\
Run V     & $5.782$ &  $-9.443$  & $100$  &  $140$  &  $0.229$ &  33000*mv1   \\
Run VISC  & $5.894$ &  $-9.495$  & $98$   &  $136$  &  $0.214$ &  30120*visc \\
\enddata
\tablecomments{The final mass of all models is $95.2\,\ME$.
Mean radius and temperature at $1\,\mathrm{bar}$ come from the
\href{https://nssdc.gsfc.nasa.gov/planetary/factsheet/saturnfact.html}{NASA Space Science Data Coordinated Archive}.
The intrinsic luminosity is from \citet{wang2024}; $L_{\sun}=3.828\times 10^{33}\,\mathrm{erg/s}$.
The effective temperature is computed from the internal heat flux estimated by
\citet{wang2024} and the equilibrium temperature (see section~\ref{sec:bound}).
}
\end{deluxetable*}

The formation and evolution of Saturn is followed through numerical simulations.
The calculations start with an $0.5\,\ME$ core of heavy elements, continue 
through the phases of concurrent accretion of $100\,\mathrm{km}$ planetesimals 
and nebular gas up to Saturn's mass, then follow the isolated phase of cooling 
and contraction with constant mass up to the time 
$t=4.57\times 10^9\,\mathrm{yr}$. 
It is no longer assumed 
\citepalias[as it was in most of the models of][and in most other 
formation/evolution models]{pollack1996}
that all accreted heavy elements sink to a condensed core which is overlaid 
by a gaseous envelope consisting mainly of H and He, with primordial nebular
composition. Here, the planetesimal material does not sink to the core. 
The dissolution and possible rain-out of the planetesimals,
composed of silicates, iron,  and ice, in the gaseous envelope are included.
Helium is insoluble in a portion of the metallic hydrogen region of the
interior during the final phase of cooling and contraction. 
The rain-out of this element contributes significantly to the luminosity 
of the planet.

The formation process is divided into up to five phases. The standard case,
summarized here, has four phases (see Table~\ref{table:smodel}). In Phase~I, 
accretion is dominated by heavy elements. The phase starts at 
$t=6.0 \times 10^5\,\mathrm{yr}$ and ends at 
$t=8.77 \times 10^5\,\mathrm{yr}$, for an elapsed time of 
$2.77 \times 10^5\,\mathrm{yr}$. 
The mass of the planet is $M_p=12.2\,\ME$ at the end of Phase~I. In Phase~II,
gas accretion dominates solid accretion by a factor of a few. The phase ends 
at $t=2.69 \times 10^6\,\mathrm{yr}$ and has an elapsed time of 
$1.81 \times 10^6\,\mathrm{yr}$. 
The end of the phase, known as crossover, has $M_Z=M_{XY}=16.2\,\ME$. 
In Phase~III, which does not exist if the protoplanetary disk has 
a small enough viscosity, runaway gas accretion takes place over 
an interval of $\sim~10^5\,\mathrm{yr}$, with a maximum 
$\dot{M}_{XY} + \dot{M}_Z = 1.06 \times 10^{-3}\,\ME\,\mathrm{yr}^{-1}$.
At the end of the phase, the planet mass is $M_p=53.5\,\ME$.
Phase~IV has an elapsed time of $8 \times 10^4\,\mathrm{yr}$ and 
is characterized by rapid contraction, detachment from the disk, and 
disk-limited accretion of gas and solids. The mass $M_p$ increases to
$95.2\,\ME$ at $t=2.87 \times 10^6\,\mathrm{yr}$,
of which $20.33\,\ME$ is in heavy elements. 
The corresponding elapsed times for Jupiter \citep{stevenson2022} during 
the four phases are $0.14$, $2.66$, $0.11$, and $0.13$, in units of
$10^6\,\mathrm{yr}$.
After Phase~IV, the planet evolves at constant mass, with gradual helium 
separation, through the phase of cooling and contraction, up to the present
time of $4.57 \times 10^9\,\mathrm{yr}$. 
The characteristics of 
the formation and evolution in this standard case are summarized in 
Table~\ref{table:smodel}. The results at the present time are shown in 
Figures~\ref{fig:5} and \ref{fig:6}.

The end result is a planet with a central region of $100$\% heavy elements 
with a mass of $12.4\,\ME$ and a radius of about $22$\% of Saturn's total
radius ($2.0\,\RE$). These results vary only slightly from case to case.
 These cases include
(\textit{i}) a run without helium separation, 
(\textit{ii}) a run (P2) in which 
the parameters in the H/He phase diagram are modified, 
(\textit{iii}), a run (V) in which $2.2\,\ME$ of 
solid material are added after gas accretion stops,
and (\textit{iv}) 
a run (VISC) in which the viscosity in the protoplanetary disk is reduced 
during disk-limited accretion. The total
heavy element mass in the planet is $M_Z=20.33\,\ME$ in all cases, except 
in Run~V, where $M_Z=22.55\,\ME$. 
Three further runs are presented, which start at the same state as the
standard case but do not allow for a comparison with the present planet
at $t=4.57 \times 10^9$ years. The first and second assume initial reduced
solid surface densities of $2 .$0 and $2.7\,\mathrm{g\,cm^{-2}}$, respectively,
and the third assumes a planetesimal size of $5\,\mathrm{km}$ 
with standard initial solid surface density.

In the outer core, energy transport is by convection. Outside this core is 
a layer of steadily decreasing mass fraction of heavy elements, containing
about $8\,\ME$ of heavy elements. In this layer, convection is suppressed, 
and radiation provides the energy transport.
The layer with a composition gradient ($dZ/dr < 0$) extends outward to $63$\%
of the total planet radius, $R_p$. Embedded in the volume with a $Z$-gradient
is a region ($44.6$ to $55.0$\% of Saturn's radius) with constant $Z=0.093$ 
and constant helium mass fraction, representing layers of enhanced He 
abundance derived from rain-out. 
The rain region itself extends from $55$\% to $62$\% of $R_p$ and includes 
a stable He gradient.  
Above it is a convective, adiabatic layer of uniform composition mostly
consisting of H and He in which the helium mass fraction has been reduced 
from its primordial value through mixture with material from the upper part 
of the rain-out zone. 

At the present time, the model mass fraction of He in the outer envelope 
is $Y_1 \approx 0.20$, reduced from the primordial value of $0.273$. 
The extent of the computed dilute core agrees with that deduced from 
gravity measurements and ring seismology \citep{mankovich2021}.
The heavy-element concentration in our models in the central regions is 
generally higher than that in static structure models developed to fit 
those measurements. Our standard model is in good agreement 
with Saturn's radius and intrinsic luminosity. If helium separation is not
included, the computed luminosity and radius are somewhat smaller than  
observed. A summary of results from the various models presented herein, 
at the current age, is given in Table~\ref{table:modelrecap}. Note that 
the uncertainty in the observed intrinsic luminosity is 
$\pm 0.03$ in $\log{(L/L_{\sun})}$.

These results are for specific models that include many physical processes
and provide reasonable matches for some observations, but they are not unique
in doing so.
The results, in particular those for $Y_1$, are tentative because of the
following difficulties and uncertainties. 
(\textit{i}) The computed mass fraction of heavy elements in the outer layers, 
about $3$\%, is somewhat low compared with observations. Including continued
accretion of solids after the gas accretion has terminated 
(Phase~V, see section~\ref{sec:param}) increases this number. However, 
the $Z$-value in Saturn's atmosphere is not really known. The main observed
quantity is methane whose mole fraction is $(4.7 \pm 0.2) \times 10^{-3}$
\citep[][]{fletcher2009}.
(\textit{ii}) There are many computed models for current Saturn 
\citep[e.g.,][]{mankovich2021,nettelmann2021} in which $Z$ at the center of 
the planet is significantly less than $1$. If later work shows that one of 
those is correct, then there is not agreement with our model, which has $Z=1$ 
at the center. More importantly, the mass of our $Z=1$ regions is typically 
$12\,\ME$, which is high compared with most recent static structure models
of Saturn (but see discussion in Section~\ref{sec:tesimal}).
(\textit{iii}) The assumed initial solid surface density of heavy elements 
in the disk, $\sigma= 3\,\mathrm{g\,cm}^{-2}$, compared with that assumed 
for Jupiter, $10\,\mathrm{g\,cm}^{-2}$, by \citet{stevenson2022}, implies that 
$\sigma \propto a^{-2}$. Observed values in some protoplanetary disks 
(which probe dust) appear to suggest a slower decline 
\citep[e.g.,][]{rosenfeld2013,liu2018}, but these observations do not 
provide information as far in as $\approx 10\,\mathrm{AU}$ in a disk. 
(\textit{iv}) 
The models assume spherical symmetry (i.e., they do not include rotation)
and, thus, the calculations of gravitational moments $J_2$ and $J_4$ are not
available. It is true that these moments are closely related to the moment 
of inertia factor (or normalized moment of inertia), which we can calculate
(based on a mean volumetric radius).
The results are higher than the estimated value of 
$0.22$ 
\citep[using the equatorial radius for normalization,][]{militzer2023}
by about $8$\%. But Saturn is not spherical, and rotation is important 
\cite[][e.g., estimated a volumetric to equatorial radius ratio of $0.965$]{militzer2023}.
A correction of this kind would reduce the discrepancy.
Clearly, more detailed models in the future are needed to resolve this
difficulty.
(\textit{v}) The temperature gradient in regions with composition gradients 
is uncertain. 
Double diffusive convection \citep[e.g.,][]{leconte2012} is not included as 
a transport mechanism. It is considered beyond the scope of the present
calculation, especially when one considers time scales of billions of years,
and it has its own uncertainties.
(\textit{vi}) 
The results depend on the assumed planetesimal size. Test calculations 
(Section~\ref{sec:tests}) with a size of $5\,\mathrm{km}$ show that 
the solid accretion rate in Phase~I is five times faster than that in 
the case of $100\,\mathrm{km}$  planetesimals. Use of small solids also
affects the opacity, but the rain-out of silicates in the early stages 
of accretion is not affected. The expected accretion time to 
a mass of $95.2\,\ME$ is far shorter than the typical nebular lifetime.
(\textit{vii}) As often mentioned, the equation of state at high pressures 
is uncertain. The biggest concern at present is actually the EoS at 
pressures of a few to many tens of kilobars, where interpolations are 
typically used to connect experiments and quantum mechanical results.
(\textit{viii}) The accretion rates for planetesimals are uncertain; 
accretion models for Jupiter using a detailed multi-zone approach and 
a range of planetesimal sizes \citep{gennaro2021}, give rates quite 
different from those derived by the methods assumed here; no such 
calculations are available for Saturn. 
Their simulations show that a planetesimal size of about 
$50\,\mathrm{km}$ gives the highest solid accretion rate.
(\textit{ix}) The accretion time for Saturn, $2.87\,\mathrm{Myr}$, is short
compared with that of Jupiter, $3.2\,\mathrm{Myr}$, computed with similar
physics \citep{stevenson2022}. The main difference occurs during Phase~II.
A small ($10$\%) reduction in the assumed initial solid surface density
(Section \ref{sec:tests}) shows that the formation time increases to
$3.8\,\mathrm{Myr}$; thus an even smaller reduction in the initial 
$\sigma$ could still result in a reasonable formation time.
(\textit{x}) There is no direct measurement, via atmospheric probe, 
of the helium mass fraction $Y_1$ in Saturn's outer layers.
(\textit{xi}) As emphasized in Section~\ref{sec:he}, the phase diagram in 
the He rain region is not well known. Uncertainties include the pressures 
in that region, the rapidity with which immiscibility grows with pressure 
(i.e., $\beta$ in Equation~(\ref{eq:6})), and the fluid dynamics of 
the He rain region.  

\begin{acknowledgments}
Primary funding for this project was provided by NASA's Research Opportunities 
in Space and Earth Science (ROSES) grants EW22-0029 and EW18-2-0060.
G.D.\ acknowledges support from NASA's ROSES grants 80HQTR21T0027 and 
NNH230B139A.
We are indebted to Ravit Helled, who provided the equation of state tables. 
We thank Jacob Kegerreis for comments on the manuscript.
\end{acknowledgments}




\end{document}